\documentclass[superscriptaddress,rmp]{revtex4}
\usepackage{amsmath}
\usepackage{amsfonts}
\usepackage{amssymb}
\usepackage{color}
\usepackage{epsfig}
\usepackage{verbatim}
\usepackage{eurosym}

\renewcommand{\a}{\alpha}
\renewcommand{\b}{\beta}
\renewcommand{\d}{\delta}
\newcommand{\g}{\gamma}
\newcommand{\s}{\sigma}

\newcommand{\e}{\epsilon}

\newcommand{\w}{\omega}
\renewcommand{\u}{\upsilon}
\renewcommand{\l}{\lambda}
\renewcommand{\k}{\kappa}
\renewcommand{\L}{\Lambda}

\newcommand{\sq}{\sigma^2}
\newcommand{\vq}{\varsigma^2}

\newcommand{\ud}{\mathrm{d}}
\renewcommand{\P}{\mathcal{P}}
\newcommand{\E}{\mathbb{E}}

\newcommand{\N}{\mathcal{N}}
\newcommand{\W}{\mathcal{W}}
\newcommand{\F}{\mathcal{F}}
\newcommand{\R}{\mathbb{R}}
\newcommand{\beq}{\begin{equation}}
\newcommand{\eeq}{\end{equation}}
\newcommand{\bea}{\begin{eqnarray}}
\newcommand{\eea}{\end{eqnarray}}

\begin{document}

\title{Informations in models of evolutionary dynamics}

\author{Olivier Rivoire} 
\affiliation{CNRS, LIPhy, F-38000 Grenoble, France.}
\affiliation{Univ.~Grenoble Alpes, LIPhy, F-38000 Grenoble, France.}


\begin{abstract}
Biological organisms adapt to changes by processing informations from different sources, most notably from their ancestors and from their environment. We review an approach to quantify these informations by analyzing mathematical models of evolutionary dynamics, and show how explicit results are obtained for a solvable subclass of these models. In several limits, the results coincide with those obtained in studies of information processing for communication, gambling or thermodynamics. In the most general case, however, information processing by biological populations shows unique features that motivate the analysis of specific models.
\end{abstract}

\maketitle

\section{Introduction}

Concepts from information theory are central to many quantitative studies of information processing in biology~\cite{Bialek}. In particular, the mutual information is commonly used to analyze input-output relationships in cellular processes such as biochemical sensing and transcriptional regulation~\cite{Nemenman:2011tb,Tkacik:2011jr,Brennan:2012uo,Bowsher:2014cd}. As a generic measure of information transmission, the mutual information has indeed a number of attractive mathematical properties~\cite{CoverThomas91}. As a measure of biological information, however, it has several shortcomings: it does not account for the organization of cells into populations or for the role of inherited information and, more generally, its connection to evolutionary fitness may be questioned. How should the mutual information be amended to account for these features? Are such amendments always decreasing the value of information, thus conferring to the mutual information the role of an ``ideal'' upper bound? Or can these amendments have a major incidence on the way information is optimally processed by a cell?\\ 

A principled approach to these questions is to follow Shannon's example~\cite{Shannon:1948wk} in defining and studying an abstract mathematical model that captures the essence of the problem of interest without directly (or axiomatically) prescribing a formula for quantifying information. This formula is instead expected to emerge as a property of the model. We review here such an approach to the problem of formalizing information processing in growing populations~\cite{Rivoire:2011ue}. Because of similarities but also differences with engineering problems, this approach leads to measures of informations that are related but not identical to those obtained from models of communication.\\

One crucial difference is that cells reproduce and form populations. This feature is common to problems of gambling and financial investment. The first analysis of the value of information in growing populations was in fact performed by Kelly in relation to horse-race gambling~\cite{KellyJr:1956un}. He found that the mutual information emerges from the analysis of his model as it does from Shannon's model of communication~\cite{Shannon:1948wk}. His results were later extended to show that, in more general models, the mutual information provides only an upper bound on the value of information~\cite{Barron:1988ul,CoverThomas91}. Several studies have pointed out the relevance of these results to biological populations~\cite{Bergstrom:2004um,Kussell:2005tk,Taylor:2007uc}. In one of them~\cite{Rivoire:2011ue}, we analyzed two other generic limitations of the mutual information as a measure of the value of biological information: its failure to account for constraints of causality, which has also been examined in the context of gambling~\cite{Permuter:2011jr}, and its failure to account for the distributed nature of biological information processing, where each individual cell processes its own information, which has no equivalent in gambling. This second feature implies that the value of information may exceed the value given by the mutual information~\cite{Rivoire:2011ue,Cheong:2011jp}.\\

Practically, deriving measures of information from abstract models is limited by the difficulty of analyzing mathematically models of sufficient generality. We show here how explicit formulae for the values of acquired and inherited informations in growing populations can be obtained for a class of solvable Gaussian models~\cite{Rivoire:2014kta}. Gaussian approximations are common in studies of information processing by biochemical networks~\cite{Ziv:2007cb,Tkacik:2008dq,Tostevin:2009ja,Cheong:2011jp}. Gaussian models of population dynamics have also their counterpart in several other fields. In information theory, they correspond to models of transmission of continuous signals in presence of additive white Gaussian noise~\cite{Shannon:1949uo}. In population genetics, Gaussian models are at the foundation of quantitative genetics, which studies the evolution of continuous traits~\cite{Lynch98}. In stochastic control theory, they are related to the Kalman filter, a tracking algorithm based on noisy measurements~\cite{Kalman:1960tn}. In physics finally, we shall present a formal mapping to the problem of controlling by feedback a Brownian particle in a tunable harmonic potential.\\

A more general connection between measures of information in growing populations and in stochastic thermodynamics was presented recently by Vinkler, Permuter and Merhav~\cite{Vinkler:2014vc}. Quantifying the value of information for controlling thermodynamical systems has been the object of many studies~\cite{Parrondo:2015cv}. Most of them follow the approach advocated here: a model is defined based on thermodynamical principles and a measure for the value of information is inferred from an analysis of its physical properties; for instance, this value is identified with the maximal work that can extracted based on microscopic measurements~\cite{Parrondo:2015cv}. Given the different premises, it is all the more interesting to find that analogous formulae emerge when analyzing information processing in evolutionary dynamics and thermodynamics.\\

The present work thus aims at connecting and extending different lines of work. In the first part, we review the problem of quantifying informations in a discrete model of growing population~\cite{Rivoire:2011ue}. Several aspects are common between this problem in gambling and in biology and we highlight the features specific to biological populations. In a second part, we show how this model becomes analytically solvable in a continuous limit. The Gaussian model thus defined extends a model studied by Haccou and Iwasa~\cite{Haccou:1995tf} and can itself be extended to a more general model~\cite{Rivoire:2014kta}. In a third part, we present and develop an analogy to problems of stochastic thermodynamics~\cite{Vinkler:2014vc}, which we apply to Gaussian models. Finally, we conclude by discussing some open challenges.

\section{Discrete model}

We start by reviewing the properties of a discrete model of information processing in growing populations~\cite{Rivoire:2011ue}.

\subsection{Definition}

\begin{figure}
\centering
\includegraphics[width=.75\linewidth]{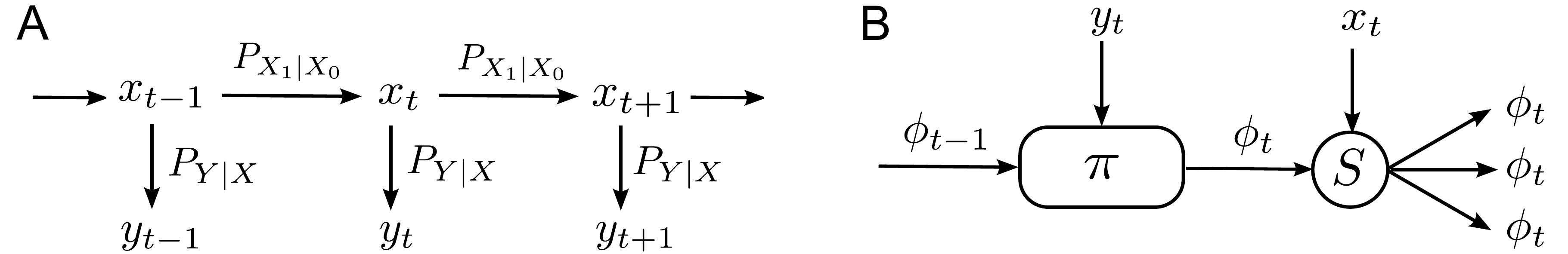}
\caption{Discrete model -- {\bf A.} The environment is a stochastic process with two components: a selective pressure $x_t$ and a cue $y_t$. The selective pressure $x_t$ follows a Markov process with conditional probability $P_{X_1|X_0}(x_t|x_{t-1})$ and the cue $y_t$ derives from $x_t$ with conditional probability $P_{Y|X}(y_t|x_t)$. {\bf B.} A member of the population at generation $t$ receives two informations, an inherited type $\phi_{t-1}$, which may differ from individual to individual, and an environmental cue $y_t$, which is common to all individuals of the same generation $t$. From these two informations, the type $\phi_t$ is generated with conditional probability $\pi(\phi_t|\phi_{t-1},y_t)$. The fitness of $\phi_t$ given the selective pressure $x_t$ decides the number $\xi$ of descendants of the individual, with $S(\phi_t,x_t)$ representing the mean value of $\xi$ given $\phi_t$ and $x_t$. The descendants inherit the type $\phi_t$ of their ancestor and are themselves subject to the next environment $(x_{t+1},y_{t+1})$. At any given time, the composition of the population is characterized by the number $N_t(\phi_t)$ of individuals of each type $\phi_t$.\label{fig:model}}
\end{figure}

The model considers a population of non-interacting individuals reproducing asexually in an independently varying environment (Figure~\ref{fig:model}). This environment is characterized by a state $x_t$, whose dependency on past history $x^{t-1}=(x_1,\dots,x_{t-1})$ is represented by a conditional probability $P_{X_t|X^{t-1}}(x_t|x^{t-1})$ (we follow the convention of denoting random variables by upper-cases and values that they take by lower-cases). An individual at generation $t$ is characterized by an internal discrete state, $\phi_t$, called its ``type'', which determines its reproductive success. This reproductive success is quantified by $S(\phi,x_t)$, the expected number of descendants in the following generation, given the internal state $\phi_t$ and the external state $x_t$. If $R(\xi|\phi_t,x_t)$ is the probability for an individual of type $\phi_t$ and in environment $x_t$ to have $\xi$ descendants in the next generation (including itself) this reproductive success is thus given by $S(\phi_t,x_t)=\langle\xi\rangle_{\phi_t,x_t}=\sum_\xi\xi R(\xi|\phi_t,x_t)$.\\

The type $\phi_t$ of an individual may depend on two things: the type $\phi_{t-1}$ of its parent and a cue $y_t$ correlated to the selective pressure $x_t$ by a conditional probability $P_{Y_t|X_t}(y_t|x_t)$, which we assume to be fixed: $P_{Y_t|X_t}(y_t|x_t)=P_{Y_1|X_1}(y_t|x_t)$ [also abbreviated $P_{Y|X}(y_t|x_t)$]. The ancestral type $\phi_{t-1}$ represents an inherited information and the perceived signal $y_t$ an acquired information. The relationship between $\phi_t$, $\phi_{t-1}$ and $y_t$ is generally considered to be stochastic, and characterized by a conditional probability $\pi(\phi_t|\phi_{t-1},y_t)$. This conditional probability $\pi$ encodes the information processing strategy followed by each individual of a population, each having its own $\phi_{t-1}$ and $\phi_t$ but experiencing the same $x_t$ and $y_t$.\\

While the model can be studied more generally~\cite{Rivoire:2011ue}, we analyze it here under two simplifying assumptions:

\ (i) We assume that the environment is stationary, ergodic and Markovian, with $P_{X_t|X^{t-1}}(x_t|x^{t-1})=P_{X_1|X_0}(x_t|x_{t-1})$.

(ii) We assume that $S(\phi_t,x_t)$ is of the form
\beq\label{eq:S=KD}
S(\phi_t,x_t)=K(x_t)\Delta(x_t|\phi_t)\quad{\rm with}\quad \Delta(x_t|\phi_t)\geq 0\quad{\rm and}\quad\sum_{x_t}\Delta(x_t|\phi_t)=1.
\eeq
This assumption means that no type $\phi_t$ has a systematic advantage when considering all possible environments $x_t$~\cite{Haccou:1995tf}. (Here and below, a notation of the type $A(u|v)$ always signifies that $A$ is a transition matrix, with $A(u|v)\geq 0$ and $\sum_uA(u|v)=1$ for all $v$.) 

\subsection{Fitness and optimality}

The dynamics of the model is summarized by a recursion for $N_t(\phi_t)$, the expected number of individuals of type $\phi_t$ at generation $t$,
\beq\label{eq:basic}
N_t(\phi_t)=\sum_{\phi_{t-1}}S(\phi_t,x_t)\pi(\phi_t|\phi_{t-1},y_t)N_{t-1}(\phi_{t-1}),
\eeq
where the series of environmental states  $x^t=(x_1,\dots,x_t)$ and cues $y^t=(y_1,\dots,y_t)$ are considered as externally fixed.\\

Quantifying the values of the inherited information $\phi_{t-1}$ and acquired information $y_t$ requires a well-defined fitness function. This fitness function should indicate the outcome of natural selection when two populations with different strategies $\pi_1$ and $\pi_2$, defining two ``species'', are competing. As this outcome may be stochastic, such a fitness function need not exist (or may depend on the particular realization of the stochastic processes). For our simple model, however, a population will, in the long term, either become extinct or grow exponentially. In the second case, the rate of exponential growth, $\L$, depends on the strategy $\pi$, the selection $S$, and the environmental parameters $P_{X_1|X_0}$ and $P_{Y_1|X_1}$, but not on the particular realization of the dynamics [mathematical details may be found in~\cite{Rivoire:2011ue}]. This growth rate thus defines a fitness function to compare the long-term value of different strategies $\pi$.\\ 

More precisely, the growth rate is given by the limit
\beq\label{eq:Lbda1}
\L=\lim_{t\to\infty}\frac{1}{t}\ln\frac{N_t}{N_0},
\eeq
where $N_t=\sum_{\phi_t}N_t(\phi_t)$ represents the expected total population size at generation $t$. If the environment is stationary and ergodic, which we shall assume, $\L$ can also be written as 
\beq\label{eq:Lbda2}
\L=\E[\ln W_t],
\eeq
where $W_t=N_t/N_{t-1}$ represents the factor by which the population size is multiplied between two successive generations, and $\E$ is an expectation with respect to the external random variables $X^t$ and $Y^t$.\\

$\L(\pi)$ defines a relevant measure of fitness in the sense that, in the long run ($t\to\infty$) and all other things being equal, a population following strategy $\pi_1$ will almost surely exponentially out-number a population following $\pi_2$ if and only if $\L(\pi_1)>\L(\pi_2)$ (provided the population does not become extinct). An optimal strategy $\hat\pi$ can therefore be defined as a strategy optimizing $\L(\pi)$. 

\subsection{Informations}

We define the value of an information as the increment of fitness that it may confer. This involves a comparison between the growth rate of two models, one in which the information is available, and one in which it is not. Mathematically, no information can be acquired when $\pi$ is of the form $\pi(\phi_t|\phi_{t-1})$ and no information is inherited when it is of the form $\pi(\phi_t|y_t)$. More generally, let $\P_0$ be a subset of the set $\P_1$ of admissible strategies in which $\pi$ is prevented from accessing a particular information. Then we define the value of this information as
\beq\label{eq:P0P1}
I = \max_{\pi\in\P_1}\L(\pi)-\max_{\pi\in\P_0}\L(\pi).
\eeq
In particular, the value of acquired information $I_{\rm acquired}$ is defined by considering the subset $\P_0$ of strategies of the form $\pi(\phi_t|\phi_{t-1})$, and the value of inherited information $I_{\rm inherited}$ of the form $\pi(\phi_t|y_t)$. By taking for $\P_0$ the subset of strategies of the form $\pi(\phi_t)$, we also define the joint value of the two informations, $I_{\rm tot}$, which is generally {\it not} the sum $I_{\rm acquired}+I_{\rm inherited}$, since $I_{\rm tot} = \max_{\pi(\phi_t|\phi_{t-1},y_t)}\L(\pi)- \max_{\pi(\phi_t)}\L(\pi)\neq \max_{\pi(\phi_t|\phi_{t-1},y_t)}\L(\pi)-\max_{\pi(\phi_t|\phi_{t-1})}\L(\pi)+ \max_{\pi(\phi_t|\phi_{t-1},y_t)}\L(\pi)-\max_{\pi(\phi_t|y_t)}\L(\pi)$.\\

\begin{figure}
\centering
\includegraphics[width=.8\linewidth]{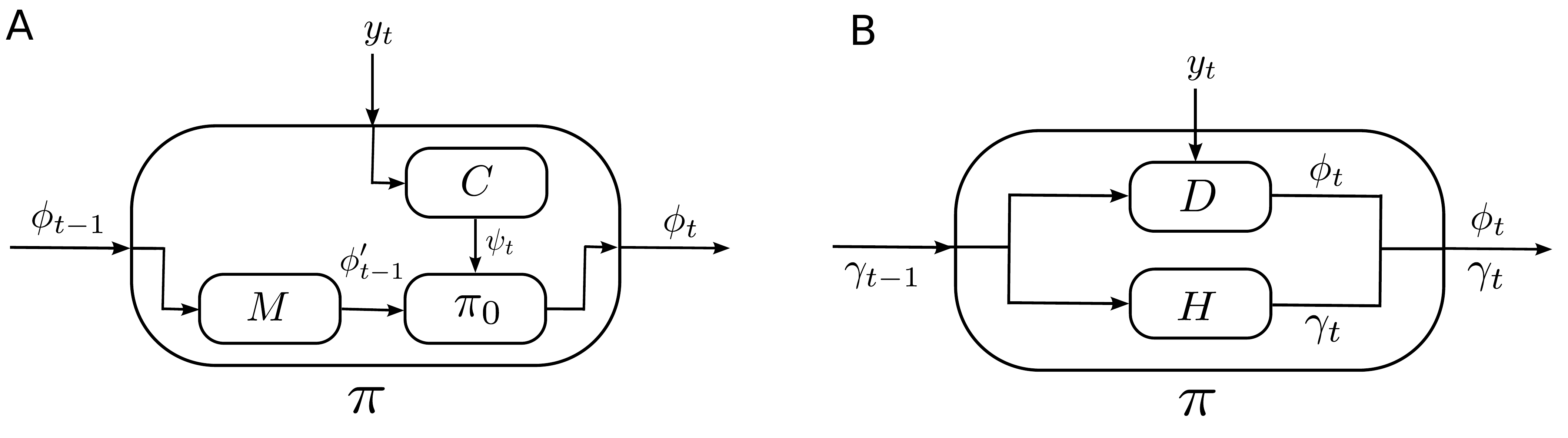}
\caption{Constrained information processing -- The conditional probability $\pi(\phi_t|\phi_{t-1},y_t)$ which decides the type $\phi_t$ of an individual given the inherited type $\phi_{t-1}$ and the environmental cue $y_t$ (Figure~\ref{fig:model}) may be constrained. {\bf A.} Replication may be subject to mutations such that the individual effectively inherits $\phi'_{t-1}$ with conditional probability $M(\phi_{t-1}'|\phi_{t-1})$. Sensing may be subject to noise such that the individual effectively perceives $\psi_t$ with conditional probability $C(\psi_t|y_t)$. Given these two elements, an individual generates its type $\phi_t$ with conditional probability $\pi_0(\phi_t|\phi'_{t-1},\psi_t)$. While the environmental cue $y_t$ is common to all members of the population at a given generation $t$, the perceived signal $\psi_t$ may differ from individual to individual.  {\bf B.} The type may have two components: a phenotype $\phi_t$ which decides the number of descendants via $S(\phi_t,x_t)$ and a genotype $\g_t$ which defines the information transmitted to these descendants. The first may be described by a conditional probability $H(\g_t|\g_{t-1})$ and the second by $D(\phi_t|\g_{t-1},y_t)$, where $\g_{t-1}$ represent the inherited genotype, which, by definition, is the only component of the type $(\phi_{t-1},\g_{t-1})$ to be inherited. \label{fig:constraints}}
\end{figure}

Additional constraints may be present that restrain $\P_0$ and $\P_1$ to a subclass of admissible strategies. For instance, the transmission of inherited information may be noisy because of random mutations following replication, with $\pi$ necessarily of the form $\pi(\phi_t|\phi_{t-1},y_{t-1})=\sum_{\phi'_{t-1}}\pi_0(\phi_t|\phi'_{t-1},y_t)M(\phi_{t-1}'|\phi_{t-1})$, where $M(\phi_{t-1}'|\phi_{t-1})$ is a given mutational matrix, and where only the conditional probability $\pi_0(\phi_t|\phi_{t-1}',y_t)$ is subject to optimization (Figure~\ref{fig:constraints}A). This corresponds to replacing Eq.~\eqref{eq:basic} by $N_{t}(\phi_{t})=\sum_{\phi'_{t-1}}S(\phi_t,x_t)\pi_0(\phi_t|\phi_{t-1}',y_t)N'_{t-1}(\phi_{t-1}')$ where $N_{t-1}(\phi_{t-1}')=\sum_{\phi_{t-1}}M(\phi'_{t-1}|\phi_{t-1})N_{t-1}(\phi_{t-1})$ represents the number of individuals mutated to $\phi'_{t-1}$.\\

Similarly, the acquisition of an information from the environmental variable $y_t$ may be limited by a noisy sensor $C(\psi_t|y_t)$, with $\pi$ constrained to be of the form $\pi(\phi_t|\phi_{t-1},y_t)=\sum_{\psi_t}\pi_0(\phi_t|\phi_{t-1},\psi_t)C(\psi_t|y_t)$ (Figure~\ref{fig:constraints}A). This constraints introduces a distinction between two types of informations: $y_t$, which is a feature of the environment and is common to all members of the population at generation $t$, and $\psi_t$, which is associated with a particular individual (we use Roman letters for environmental variables and Greek letters for individual variables). For instance, $y_t$ may represent the concentration of one of several constituents of the environment, related to $x_t$ by $P_{Y|X}(y_t|x_t)$, and $\psi_t$ the concentration of this constituent as perceived by a particular individual, given its noisy sensor $C(\psi_t|y_t)$. The cue $y_t$ and the sensor $C$ are common to all individuals but not necessarily the perceived signal $\psi_t$. This decomposition may be viewed as the counterpart at a population level of the decomposition between extrinsic and intrinsic noise at the individual level~\cite{Swain:2002ww}: as intrinsic noise corresponds to intra-individual variations and extrinsic noise to inter-individual variations in gene expression, the intrinsic information $\psi_t$ corresponds to intra-generation variations and the extrinsic information $y_t$ to inter-generation variations in information sensing. This distinction becomes important when evaluating the value of the information provided by the sensor $C(\psi_t|y_t)$, as opposed to the value of the information provided by the ``environmental channel'' $P_{X|Y}(y_t|x_t)$ (see examples below).\\

Another biologically motivated constraint on $\pi$ is the decomposition of the type of an individual into a genotype, which is inherited and transmitted, and a phenotype on which selection acts. A generic model making this distinction is for instance defined by the recursion
\beq\label{eq:gp}
N_t(\g_t)=\sum_{\phi_t,\g_{t-1}}\ S(\phi_t,x_t) H(\g_t|\g_{t-1},\phi_t,z_t)D(\phi_t|\g_{t-1},y_t)N_{t-1}(\g_{t-1}),
\eeq
where $D(\phi_t|\g_{t-1},y_t)$ specifies how the phenotype $\phi_t$ stochastically depends on the inherited genotype $\g_{t-1}$ and some aspect $y_t$ of the environment and $H(\g_t|\g_{t-1},\phi_t,z_t)$ how the transmitted genotype $\g_t$ depends on the inherited genotype and some possibly different aspect $z_t$ of the environment (Figure~\ref{fig:constraints}B). As shown in Appendix~\ref{app:mapping}, this model corresponds to Eq.~\eqref{eq:basic} when $\pi$ is constrained to a particular set of admissible strategies. The model defined by Eq.~\eqref{eq:gp}, however, has two acquired informations: $y_t$ at the phenotypic level and $z_t$ at the genotypic level (which each may be decomposed into extrinsic and intrinsic contributions). This extension corresponds to a discrete version of the model proposed in~\cite{Rivoire:2014kta} and illustrates the fact that multiple acquired informations may be defined and quantified. Similarly, the model can be extended to deal with multiple inherited informations, for instance to represent a genetic and an epigenetic contribution to heredity. 

\subsection{Solvable limits}

In two limits, Eq.~\eqref{eq:basic} factorizes into a recursion that involves only the total population size $N_t=\sum_{\phi_t}N_t(\phi_t)$. The first limit is when the environment is maximally selective, so that only one type $\phi_t$, which may be defined without loss of generality as $\phi_t=x_t$, can survive in each environmental state $x_t$, 
\beq\label{eq:Kelly}
S(\phi_t,x_t)=K(x_t)\delta(x_t,\phi_t)\qquad \textrm{[perfect selectivity]}
\eeq
where $K(x_t)$ represents the multiplicative rate of the surviving type, and $\delta$ denotes the Kronecker symbol, with $\delta(x_t,\phi_t)=1$ if $\phi_t=x_t$ and 0 otherwise. This corresponds to $\Delta(x_t|\phi_t)=\delta(x_t,\phi_t)$ in Eq.~\eqref{eq:S=KD}. In this case, $N_t=N_t(x_t)$ and
\beq\label{eq:Kd}
N_t=W_tN_{t-1}\quad{\rm with}\quad W_t=K(x_t)\pi(x_t|x_{t-1},y_t).
\eeq
The second limit is in absence of inheritance, when the current type $\phi_t$ of an individual cannot depend on its ancestral type $\phi_{t-1}$,
\beq
\pi(\phi_t|\phi_{t-1},y_t)=\pi(\phi_t|y_t) \qquad \textrm{[no inheritance]}
\eeq
which implies
\beq
N_t=W_tN_{t-1}\quad{\rm with}\quad W_t=\sum_{\phi_t}S(\phi_t,x_t)\pi(\phi_t|y_t).
\eeq
Given the assumption made in Eq.~\eqref{eq:S=KD}, this may be rewritten as $N_t=K(x_t)\tilde\pi(x_t|y_t)N_{t-1}$, as in Eq.~\eqref{eq:Kd}, but with an effective strategy $\tilde\pi$ defined by
\beq\label{eq:effstrat}
\tilde\pi(x_t|y_t)=\sum_{\phi_t}\Delta(x_t|\phi_t)\pi(\phi_t|y_t).
\eeq
The effective strategy $\tilde\pi$ is here constrained to a particular subset $\P_1$, as in the examples discussed above.\\

The conjunction of the two limits, perfect selectivity and no inheritance, defines Kelly's model~\cite{KellyJr:1956un}, where
\beq\label{eq:Wkelly}
N_t=W_tN_{t-1}\quad{\rm with}\quad W_t=K(x_t)\pi(x_t|y_t),
\eeq
and therefore
\beq\label{eq:Lkelly}
\L=\E[\ln W_t]=\E_X[\ln K(X)]+\E_{X,Y}\ln [\pi(X|Y)].
\eeq
where $\E_{X}\ln K(X)=\sum_xP_{X}(x)\ln K(x)$ with $P_{X}(x)=P_{X_t}(x)$ describing the probability of $x_t=x$ (since the environment is assumed to be stationary, it is independent of $t$), and where $\E_{X,Y}[\pi(X|Y)]=\sum_{x,y}P_{X,Y}(x,y)\ln [\pi(x|y)]$ with $P_{X,Y}(x,y)=P_{Y|X}(y|x)P_X(x)$ describing the joint probability of $(x_t,y_t)=(x,y)$.\\

In the original formulation of this model~\cite{KellyJr:1956un}, $N_t$ is a capital that a gambler bets on successive horse races and $x_t$ represents the horse winning on race $t$, $K(\phi_t)$ the odds for horse $\phi_t$  (the ratio of the full payout to the stake if it wins) and $y_t$ a side-information hinting at the identity of $x_t$. The betting strategy $\pi(\phi_t|y_t)$ defines the fraction of capital bet on each horse $\phi_t$ given the information $y_t$, which the gambler wants to choose so as to maximize its cumulative gain $N_t=\prod_{k=1}^tW_kN_0$. In this interpretation, an individual corresponds to a particular unit of currency, say a 1\euro\ coin, and the ``type'' of a coin to the horse on which it is bet.\\

The analogy extends to models with finite selectivity, corresponding to multiple horses having non-zero return and to models with inheritance, corresponding to a gambler with memory~\cite{CoverThomas91}. Some aspects of information processing in biological population have, however, no analogy in gambling, such as the distinction between extrinsic and intrinsic informations. Information processing is indeed centralized in gambling, where a gambler controls each of its coins, while  it is distributed in biology, where each member of a population can act independently and stochastically.\\

In the following two sections, we summarize the properties of the model in each of the two generally solvable limits of no inheritance and perfect selectivity, before introducing a continuous model that can be solved beyond these two limits. We refer to~\cite{CoverThomas91,Permuter:2011jr,Rivoire:2011ue} for a derivation of the results.

\subsection{No inheritance}

Assuming no inheritance, i.e., $\pi$ constrained to the form $\pi(\phi_t|y_t)$, we can write the growth rate as (see Appendix~\ref{app:decamp})
\beq\label{eq:hatLtildePi}
\L=\L^*-H(X)+I(X;Y)-\E_{Y}[D(P_{X|Y}(.|Y)\| \tilde\pi(.|Y))],
\eeq
where $\tilde\pi$ is the effective strategy defined in Eq.~\eqref{eq:effstrat}. In this decomposition, each term has an interpretation of its own~\cite{CoverThomas91}:

$\bullet$ $\L^*=\E_{X}\ln K(X)=\sum_xP_{X}(x)\ln K(x)$ corresponds to a maximal growth rate, possibly achievable only if knowing exactly the sequence of environmental states;

$\bullet$ $H(X)=-\sum_x P_{X}(x)\ln P_{X}(x)$ is the entropy of $X_t$, and represents here a cost due to the stochasticity of environmental process;

$\bullet$ $I(X;Y)$ is the mutual information between the cue $Y_t$ and the selective variable $X_t$, defined by
\beq\label{eq:mi}
I(X;Y)=\sum_{x,y}P_{Y|X}(y|x)P_X(x)\ln\frac{P_{Y|X}(y|x)}{P_Y(y)},
\eeq
where $P_Y(y)=\sum_xP_{Y|X}(y|x)P_X(x)$ is the probability of $y_t=y$. It can also be written $I(X;Y)=H(Y)-H(Y|X)$ or $I(X;Y)=H(X)-H(X|Y)$ if introducing the conditional entropy $H(Y|X)=-\sum_{x,y}P_{X,Y}(x,y)\ln P_{Y|X}(y|x)$. The mutual information represents here a gain due to the information about $X_t$ that is contained in $Y_t$ and is zero if and only if $X_t$ and $Y_t$ are independent random variables;

$\bullet$ $\E_{Y}[D(P_{X|Y}(.|Y)\| \tilde\pi(.|Y))]=\sum_y P_Y(y)D(P_{X|Y}(.|y)\| \tilde\pi(.|y))$ represents the cost of following a suboptimal strategy. It involves a relative entropy, which is generally defined between two distributions $P(x)$ and $Q(x)$ as
\beq
D(P\|Q)=\sum_xP(x)\ln\frac{P(x)}{Q(x)}.
\eeq
$D(P\|Q)\geq 0$ and $D(P\|Q)= 0$ if and only if $P=Q$. It also involves $P_{X|Y}$, the conditional probability of $X_t$ given $Y_t$, which by Bayes' rule is given by
\beq\label{eq:bayesian}
P_{X|Y}(x|y)=\frac{P_{Y|X}(y|x)P_X(x)}{P_Y(y)}.
\eeq

Since $\pi$ appears only in the last term of Eq.~\eqref{eq:hatLtildePi}, which is necessarily non-negative, the optimal growth rate is
\beq\label{eq:Lhid}
\hat\L=\L^*-H(X)+I(X;Y)-\min_{\pi}\E_{Y}[D(P_{X|Y}(.|Y)\| \tilde\pi(.|Y))].
\eeq
In computing the minimum, two situations may arise. If the equation $\tilde\pi=P_{X|Y}$ has a solution in $\pi$, then this solution optimizes the growth rate by reducing to zero the relative entropy term, and $\hat\L=\L^*-H(X)+I(X;Y)$. Otherwise, $\hat\L<\L^*-H(X)+I(X;Y)$.\\

When considering the value of acquired information, the optimal growth rate in absence of information, $\hat\L=\L^*-H(X)-\min_{\pi}D(P_X\| \tilde\pi)$, must also be evaluated [minimum over $\P_0$ in Eq.~\eqref{eq:P0P1}]:
\beq\label{eq:Iacq}
I_{\rm acquired}=I(X;Y)-\min_\pi\E_Y[D(P_{X|Y}(.|Y)\| \tilde\pi(.|Y))]+\min_\pi D(P_X\| \tilde\pi)].
\eeq
Since $\tilde\pi=P_X$ has a solution whenever $\tilde\pi=P_{X|Y}$ has a solution $\hat\pi(x|y)$ [given by $\pi(x)=\sum_y\hat\pi(x|y)P_Y(y)$], three cases must be considered: (i)~$\tilde\pi=P_{X|Y}$ has a solution (implying that $\tilde\pi=P_X$ has one); (ii)~$\tilde\pi=P_X$ has a solution but not $\tilde\pi=P_{X|Y}$; (iii)~$\tilde\pi=P_X$ has no solution (implying that $\tilde\pi=P_{X|Y}$ has none). In the first case, $I_{\rm acquired}=I(X;Y)$, while in the two others $I_{\rm acquired}<I(X;Y)$, as may be proved even without assuming Eq.~\eqref{eq:S=KD} ~\cite{CoverThomas91}.\\

In any case, the value of acquired information is bounded by a mutual information, $I_{\rm acquired}\leq I(X;Y)$. This mutual information, however, is between the selective pressure $X_t$ and the cue $Y_t$, both environmental variables, and {\it not} between the input $Y_t$ and the output $\Psi_t$ of the sensor of a particular individual. The mutual information $I(\Psi;Y)$ can indeed exceed $I_{\rm acquired}$ as shown explicitly with a two-state model in~\cite{Rivoire:2011ue} and with a Gaussian model below. The value of acquired information in presence of a sensor with noise $C(\psi|y)$ is
\beq\label{eq:Iacqsensor}
I_{\rm acquired}=I(X;Y)-\min_{\pi_0}\E_Y[D(P_{X|Y}(.|Y)\|\tilde\pi\ast C(.|Y))]+\min_\pi D(P_X\|\tilde\pi),
\eeq
where $\tilde\pi\ast C(x|y)=\sum_{\phi,\psi} \Delta(x|\phi)\pi_0(\phi|\psi)C(\psi|y)$. A sensor with a given noise $C(\psi|x)$ is in fact always more valuable than an environmental channel $P_{Y|X}$ with same noise~\cite{Rivoire:2011ue}. This is most simply illustrated with a model with perfect selectivity, as described by Eq.~\eqref{eq:Lkelly}. In this case, $Y_t=X_t$ implies  $\L=\L^*+\E_X\ln\pi(X|X)$ with $\pi(x|x)=\sum_\psi\pi_0(x|\psi)C(\psi|x)=\E_{\Psi|X=x} \pi_0(x|\Psi)$ and, by concavity of the logarithm,
\beq
\L=\L^*+\E_X\ln\E_{\Psi|X} \pi_0(X|\Psi)\geq \L^*+\E_{X,\Psi}\ln\pi_0(X|\Psi).
\eeq
The right-hand side corresponds to the growth rate of a model with $\Psi_t=Y_t$, where $Y_t$ is given by $P_{Y|X}(y_t|x_t)=C(y_t|x_t)$. This inequality is analogous to the statement in statistical mechanics that the quenched free energy of a disordered system is bounded from below by the corresponding annealed free energy. It represents here the benefice of multiple distributed sensors over a single centralized sensor with same noise.

\subsection{Perfect selectivity}

In the other limit of perfect selectivity, an expression formally similar to Eq.~\eqref{eq:hatLtildePi} can be written
\beq\label{eq:LKinheritance}
\L=\L^*-H(X_1|X_0)+I(X_1;Y_1|X_0)-\E_{X_0,Y_1}[D(P_{X_1|X_0,Y_1}(.|X_0,Y_1)\| \pi(.|X_0,Y_1))],
\eeq
where a conditioning on the past environment $X_0$ needs to be added (and where $\pi$ replaces $\tilde\pi$). Here, the conditional mutual information $I(X_1;Y_1|X_0)$ is defined by $I(X_1;Y_1|X_0)=H(Y_1|X_0)-H(Y_1|X_1)$ [since $H(Y_1|X_1,X_0)=H(Y_1|X_1)$].\\

The optimum growth rate is obtained for $\pi$ minimizing the last term of Eq.~\eqref{eq:LKinheritance}. In absence of constraints, it is reached for $\hat\pi=P_{X_1|X_0,Y_1}$, corresponding to $\hat\L=\L^*-H(X_1|X_0)+I(X_1;Y_1|X_0)$. In this case, $I_{\rm acquired}=I(X_1;Y_1|X_0)$. Since $I(X_1;Y_1|X_0)=I(X_1;Y_1)-I(X_0;X_1)$, the difference with the instantaneous mutual information $I(X_1;Y_1)$, is exactly $I(X_0;Y_1)$, the value of the cue $Y_t$ that is already contained in the knowledge of the past environmental state $X_{t-1}$. More generally, with constraints, the last term may not vanish and $I_{\rm acquired}\leq I(X_1;Y_1|X_0)$.\\

The value of inherited information is read from another equivalent decomposition of the growth rate where $X_0$ and $Y_1$, which play similar roles, are formally exchanged:
\beq
\L=\L^*-H(X_1|Y_1)+I(X_1;X_0|Y_1)-\E_{X_0,Y_1}[D(P_{X_1|X_0,Y_1}(.|X_0,Y_1)\| \pi(.|X_0,Y_1))].
\eeq
This implies $I_{\rm inherited}\leq I(X_1;Y_1|X_0)$, where the conditional mutual information $I(X_0;X_1|Y_1)$ takes into account that some of the information contained in $X_{t-1}$ is also present in $Y_t$.\\

Finally, the total information conferred by the two sources of information satisfies
\beq\label{eq:Itot}
I_{\rm tot}\leq I(X_1;X_0)+I(X_1;Y_1|X_0)=I(X_1;Y_1)+I(X_0;X_1|Y_1),
\eeq
with equality in absence of constraints.\\

In presence of inheritance, the role of the mutual information is thus played by a conditional mutual information. The conditional mutual information $I(X_1;Y_1|X_0)$ not only differs from the instantaneous mutual information $I(X_1;Y_1)$, but also from the rate of path/trajectory mutual information, which is defined from the mutual information $I(X^t;Y^t)$ between the processes $X^t=(X_1,\dots,X_t)$ and $Y^t=(Y_1,\dots,Y_t)$ as $\lim_{t\to\infty}I(X^t;Y^t)/t$. The difference becomes apparent when applying the chain rule~\cite{CoverThomas91} to write $I(X^t;Y^t)=\sum_{k=1}^tI(X_k;Y^t|X_{k-1})$ since $I(X_k;Y^t|X_{k-1})\geq I(X_k;Y_k|X_{k-1})=I(X_1;Y_1|X_0)$, with, in general, a strict inequality. This inequality accounts for a constraint of causality: an individual has access at time $t$ to the present cue $y_t$, but not to future cues $y_k$ with $k>t$, which could allow for a better estimation of $x_t$ if available. These considerations extend in non-Markovian environments to strategies of the form $\pi(\phi_t|\phi^{t-1},y^t)$, where an individual has access to past cues $y_k$ with $k<t$~\cite{Permuter:2011jr}. The value of acquired information then corresponds to the more general concept of directed information, denoted $I(Y\to X)$, which appears repeatedly in problems of feedback control where constraints of causality are involved~\cite{Massey:1990vy}. The conditional mutual information $I(X_1;Y_1|X_0)$ is the particular value taken by the directed information $I(Y\to X)$ when considering stationary, Markovian stochastic processes. The directed information generally differs from the transfer entropy, also proposed to quantify the causal relationships between stochastic processes~\cite{Schreiber:2000wo,Amblard:2013gv}.

\section{Gaussian model}

We now present a continuous limit of the discrete model for which the growth rate $\L$ can be computed analytically beyond the two cases of perfect selectivity and no inheritance.

\subsection{Definition}

A model with continuous traits $\phi_t\in\R$ is defined by replacing Eq.~\eqref{eq:basic} with
\beq
n_t(\phi_t)=\frac{1}{W_t}\int\ud\phi_{t-1}\ S(\phi_t,x_t)\pi(\phi_t|\phi_{t-1},y_t)n_{t-1}(\phi_{t-1}),
\eeq
where $n_t(\phi_t)$ represents the density of individuals with trait $\phi_t$ in the current population, with $n_t(\phi_t)\geq 0$ and $\int\ud\phi_t\ n_t(\phi_t)=1$. The function $S(\phi_t,x_t)$ is chosen as in Eq.~\eqref{eq:S=KD} to be of a factorized form
\beq\label{eq:Sgauss}
S(\phi_t,x_t)=K(x_t) G_{\sq_s}(\phi_t-x_t),
\eeq
where $G_{\sq}(x)=(2\pi\sq)^{-1/2}\exp(-x^2/2\sq)$ represents a generic Gaussian function and $K(x)\geq 0$ is arbitrary. We parametrize $\pi$ as
\beq\label{eq:piG}
\pi(\phi_t|\phi_{t-1},y_t)=G_{\sq_\pi}(\phi_t-\l\phi_{t-1}-\k y_t),
\eeq
where $\sq_\pi$ quantifies the degree of stochasticity, $\l$ the contribution of the inherited information and $\k$ of the acquired information (it can be shown that the optimal $\pi$ is necessarily of this form).\\

The growth rate $\L$ associated with this model can be computed analytically for different environmental processes, but we consider here a stationary Markovian Gaussian process, i.e., a discrete Ornstein-Uhlenbeck process:
\beq\label{eq:OU}
P_{X_1|X_0}(x_{t+1}|x_t)=G_{\sq_{x_1|x_0}}(x_{t+1}-ax_t),
\eeq
where $a<1$ parametrizes the temporal correlation between successive environments and $\sq_{x_1|x_0}=\E[X_1^2|X_0]$ the amplitude of their variations. This interpretation follows from noticing that $\E[X_tX_1]=a^{2t}$ and $\sq_{x_1}=\E[X_1^2]=\sq_{x_1|x_0}/(1-a^2)$. Finally, we take a Gaussian channel for $P_{Y_1|X_1}=P_{Y|X}$:
\beq
P_{Y|X}(y_t|x_t)=G_{\sq_{y_1|x_1}}(y_t-x_t),
\eeq
where  $\sq_{y_1|x_1}=\E[Y_1^2|X_1]$ represents its noise. For independent environments ($a=0$), this model was studied in~\cite{Haccou:1995tf}.\\

The growth rate $\L$ for this model can be computed analytically (see Appendix~\ref{app:LG}):
\beq\label{eq:L}
\L=\L^*-\frac{1}{2}\ln(2\pi\sq_s)+\frac{1}{2}\ln\frac{\a}{\l}-\frac{\a}{2\l(1-\a^2)\sq_s}\left[\frac{(\l^2+(1-\k)^2)(1+a\a)-2\l(1-\k)(a+\a)}{(1-a\a)(1-a^2)}\sq_{x_1|x_0}+\k^2\sq_{y_1|x_1}\right],
\eeq
where 
\beq
\alpha=\frac{2\l}{1+\l^2+\b+((1-\l^2-\b)^2+4\b)^{1/2}},\qquad \b=\frac{\sq_\pi}{\sq_s},
\eeq
and $\L^*=\E_X\ln K(X)$. The model has seven parameters, four to describe the environment, $\sq_s$ for the selectivity of the environment, $a$ for the correlation between successive environments, $\sq_{x_1|x_0}$ for the amplitude of their fluctuations and $\sq_{y_1|x_1}$ for the (extrinsic) noise of the cue, and three to describe the strategy $\pi$: $\sq_\pi$, $\l$ and $\k$.\\

The two limits of no inheritance and perfect selectivity correspond, respectively, to the limits $\l\to 0$ and $\sq_s\to 0$. We show below how, in these limits, the growth rate of this continuous model has a decomposition similar to the decomposition of the growth rate of the discrete model. With the continuous Gaussian model, however, explicit formulae for the values of information can be obtained even when they do not coincide with a mutual information. Models with constraints and not assuming any of these limits can also be treated in this same framework~\cite{Rivoire:2014kta} [see Appendix~\ref{app:pg} for the link between this model and the model in~\cite{Rivoire:2014kta}].

\subsection{No inheritance}

In absence of inheritance ($\l=0$), Eq.~\eqref{eq:L} becomes (see Appendix~\ref{app:2form}):
\beq\label{eq:Lnoinh}
\L=\L^*-h(X)+I(X;Y)-\E_Y[D(P_{X|Y}(.|Y)\| \tilde\pi(.|Y))],
\eeq
where $\tilde\pi(x|y)=G_{\sq_s}\ast\pi\ (x|y)=\int\ud\phi\ G_{\sq_s}(\phi-x)\pi(\phi|y)=G_{\sq_s+\sq_\pi}(x-\k y)$ represents an effective strategy as in Eq.~\eqref{eq:effstrat}, and where 
\beq\label{eq:hI}
h(X)=\frac{1}{2}\ln(2\pi e\sq_{x_1}),\qquad {\rm and}\qquad  I(X;Y)=\frac{1}{2}\ln\left(1+\frac{\sq_{x_1}}{\sq_{y_1|x_1}}\right).
\eeq
The only difference with Eq.~\eqref{eq:hatLtildePi} is the presence of a differential entropy $h(X)$ instead of the entropy $H(X)$. The differential entropy is generally defined for continuous random variables as $h(X)=-\int\ud x P_X(x)\ln P_X(x)$. While the mutual information $I(X;Y)=h(X)-h(X|Y)$ corresponds to a limit of discrete mutual informations when $X$ is discretized into an increasing number of bits, the discrete entropy diverges in this limit, and the differential entropy $h(X)$ represents only the non-diverging part~\cite{CoverThomas91}. This divergence is compensated here by the divergence of $S(\phi_t,x_t)$ when $\sq_s\to 0$ [see Eq.~\eqref{eq:Sgauss}].\\

The Gaussian model has the advantage over the discrete model that the value of acquired information $I_{\rm acquired}$ given by Eq.~\eqref{eq:Iacq} can be evaluated explicitly.
If assuming that $\pi$ is not subject to any additional constraint, three cases must be distinguished~\cite{Haccou:1995tf}:

(i) if $\sq_s\leq\sq_{x|y}$, where $\sq_{x|y}=(\s_x^{-2}+\s_{y|x}^{-2})^{-1}$, the two equations $P_{X|Y}=G_{\sq_s}\ast\pi$ and $P_X=G_{\sq_s}\ast\pi$ have a solution, respectively given by $\hat\s_\pi^2=\sq_{x|y}-\sq_s$, $\hat\kappa=1/(1+\sq_{y|x}/\sq_x)$, and $\hat\s_\pi^2=\sq_x-\sq_s$, $\hat\kappa=1$; in this case,
\beq
I_{\rm acquired}=I(X;Y)=\frac{1}{2}\ln\left(1+\frac{\sq_{x}}{\sq_{y|x}}\right).
\eeq

(ii) if $\sq_{x|y}<\sq_s\leq\sq_x$, $P_X=G_{\sq_s}\ast\pi$ has a solution but not $P_{X|Y}=G_{\sq_s}\ast\pi$, and $D(P_{X|Y}\|G_{\sq_s}\ast\pi)>0$ is minimized with $\hat\s^2_\pi=0$, $\hat\kappa=1/(1+\sq_{y|x}/\sq_x)$; in this case,
\beq\label{eq:Iacqfs}
I_{\rm acquired}=I(X;Y)-D(G_{\sq_{x|y}}\|G_{\sq_s})=\frac{1}{2}\left(\ln\frac{\sq_x}{\sq_s}-\frac{\sq_{y|x}\sq_x}{(\sq_{y|x}+\sq_{x})\sq_s}+1\right).
\eeq

\begin{figure}
\centering
\includegraphics[width=.98\linewidth]{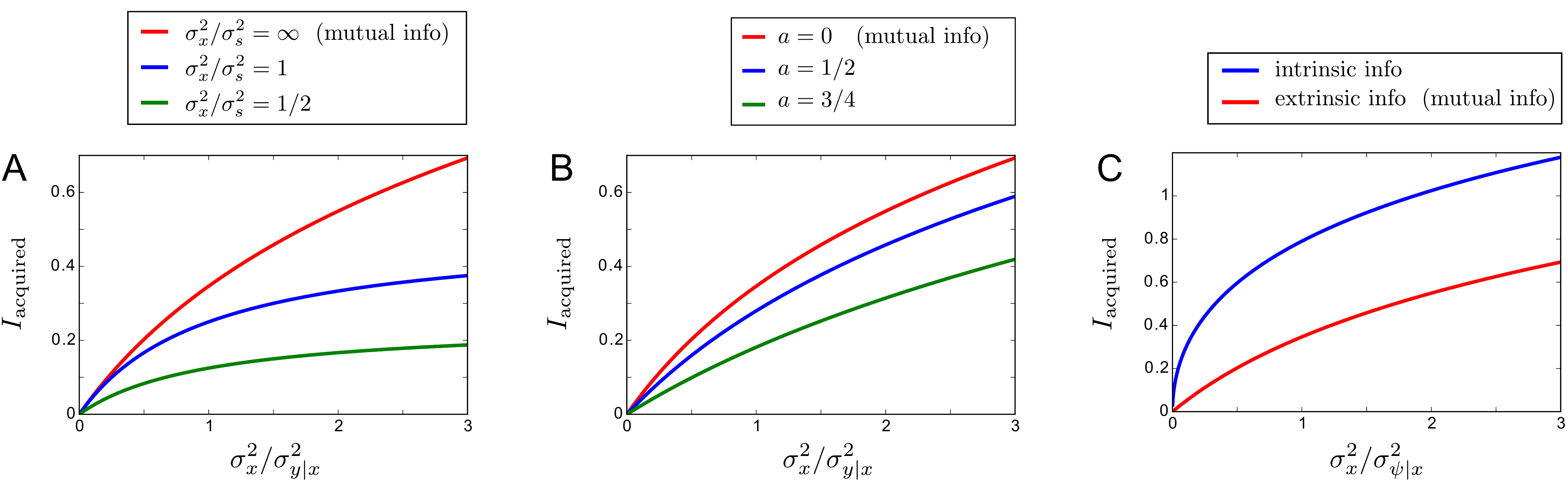}
\caption{Value of the acquired information in different limits of the Gaussian model -- {\bf A.} No inheritance but finite selectivities $\sq_s$, with $\sq_s=0$ (red curve) corresponding to the limit of perfect selectivity where the value of acquired information is given by the mutual information $I(X;Y)=(1/2)\ln(1+\sq_x/\sq_{y|x})$. A finite selectivity leads to smaller values for the acquired information (blue and green curves). Here $\sq_x=\E[X^2_t]$ represents the variance of the selective pressure $x_t$ and $\sq_{y|x}=\E[Y^2_t|X_t]$ the noise in the environmental cue $y_t$. {\bf B.} Perfect selectivity but inheritance, with $a=0$ (red curve) corresponding to the limit where inheritance has no value because the environment has no temporal correlations. In this case, and in this case only, the value of acquired information is given by the mutual information $I(X;Y)$ (red curve), otherwise it has a lower value (blue and green curves). {\bf C.}~Extrinsic versus intrinsic informations, with no inheritance and perfect selectivity, but with a possibly noisy individual sensor $C(\psi_t|y_t)$ (as in Figure~\ref{fig:constraints}A). When the sensor is noise-less ($\psi_t=y_t$) but the cue $y_t$ has a noise $\sq_{y|x}=\sq_{\psi|x}$, the value of acquired information is given by the mutual information $I(X;Y)$ (red curve, as in A and B, but note the difference of scale along the $y$-axis). When the cue is noise-less ($y_t=x_t$) but the sensor has a noise $\sq_{\psi|x}$, the value of acquired information is higher (blue curve).\label{fig:ei}}
\end{figure}

(iii) if $\sq_x<\sq_s$, neither $P_{X|Y}=G_{\sq_s}\ast\pi$ nor $P_X=G_{\sq_s}\ast\pi$ have solutions and 
\beq
I_{\rm acquired}=I(X;Y)-D(G_{\sq_{x|y}}\|G_{\sq_s})+D(G_{\sq_x}\|G_{\sq_s})=\frac{1}{2}\frac{\s^4_x}{(\sq_{y|x}+\sq_{x})\sq_s}.
\eeq
This formulae show how the value of information can depend on the degree of selectivity $\sq_s$ of the environment, in addition to the ratio signal/noise $\sq_x/\sq_{y|x}$ that controls the mutual information (Figure~\ref{fig:ei}A).\\

These different cases are associated with qualitatively different optimal strategies: (i) corresponds to an effective Bayesian strategy, $\tilde\pi=P_{X|Y}$, but (ii) and (iii) to a deterministic response, $\hat\phi_t=\hat\k y_t$, also known as a ``pure strategy'' in game theory. This later case is an example where a Bayesian inference of $x_t$ given $y_t$ is pointless: the optimal strategy is simply to act as if the information was noise-less, with only the multiplication factor $\hat\k$ to account for the presence of noise.

\subsection{Perfect selectivity}

In the limit of perfect selectivity $\sq_s\to 0$, we verify that
\beq\label{eq:Ginfsel}
\L=\L^*-h(X_1|X_0)+I(X_1;Y_1|X_0)-\E_Y[D(P_{X|Y}(.|Y)\| \pi(.|Y))],
\eeq
which is similar to Eq.~\eqref{eq:LKinheritance} but with a conditional differential entropy $h(X_1|X_0)$ instead of the entropy $H(X_1|X_0)$. We have explicitly (see Appendix~\ref{app:general}):
\beq
h(X_1|X_0)=\frac{1}{2}\ln(2\pi e\s^2_{x_1|x_0}),\qquad I(X_1;Y_1|X_0)=\frac{1}{2}\ln\left(1+\frac{\sq_{x_1|x_0}}{\sq_{y_1|x_1}}\right).
\eeq
Given Eq.~\eqref{eq:hI} and $\sq_{x_1|x_0}\leq\sq_{x_1}=\sq_{x_1|x_0}/(1-a^2)$, we verify that $I(X_1;Y_1|X_0)\leq I(X_1;Y_1)$, with a strict inequality if successive environments are non independent (Figure~\ref{fig:ei}B).\\

If $\pi$ is not constrained, the optimal strategy is $\hat\pi=P_{X_1|X_0,Y_1}$, which corresponds to (see Appendix~\ref{app:general}):
\beq\label{eq:Kgain}
\hat\k=\frac{1}{1+\sq_{y_1|x_1}/\sq_{x_1|x_0}},\qquad\hat\l=a(1-\hat\k),\qquad\hat\s_\pi^2=\hat\k\sq_{x_1|x_0}.
\eeq
While the value of acquired information is determined by $I(X_1;Y_1|X_0)$, the value of inherited information is determined by
\beq
I(X_1;X_0|Y_1)=\frac{1}{2}\ln\left(1+\frac{\sq_{x_1|y_1}}{\sq_{x_0|x_1}}\right)=\frac{1}{2}\ln\left(1+a^2\frac{\sq_{y_1|x_1}}{\sq_{x_1|x_0}}\right).
\eeq
Finally, the total value of the two informations, given in Eq.~\eqref{eq:Itot}, is at most $I(X_1;Y_1|X_0)+I(X_0;X_1)$, i.e.,
\beq
I_{\rm tot}=\frac{1}{2}\ln\left(1+\frac{\sq_{x_1|x_0}}{\sq_{y_1|x_1}}\right)+\frac{1}{2}\ln\left(\frac{1}{1-a^2}\right).
\eeq
This formulae show how the value of acquired information depends on the presence of inherited information when the successive environment are correlated ($a>0$).

\subsection{Common and individual informations}

The formulae presented so far assume the absence of constraint on $\pi$. They have to be corrected in presence of a noisy individual sensor, as shown by Eq.~\eqref{eq:Iacqsensor} in absence of inheritance. To illustrate this case in the simplest setting, we assume here both an absence of inheritance ($\l=0$) and an perfect selectivity ($\sq_s=0$), in which case Eq.~\eqref{eq:Iacqsensor} becomes
\beq\label{eq:IacqsensorG}
I_{\rm acquired}=I(X;Y)-\min_{\sq_\pi,\k}\E_Y[D(P_{X|Y}(.|Y)\|G_{\sq_\pi+\k^2 \sq_{\psi|y}}(.-\k Y))]
\eeq
since $\min_\pi D(P_X\|\pi)=0$ with $\hat\pi=P_X$. Two cases must be distinguished:

(i) if $\k^2_0\sq_{\psi|y}\leq\sq_{x|y}$, where $\kappa_0=1/(1+\sq_{y|x}/\sq_x)$ and $\sq_{x|y}=\k_0\sq_{y|x}$, the equation $P_{X|Y}(x|y)=\pi\ast G_{\sq_{\psi|y}}(x-\k y)$ has a solution given by $\hat\s_\pi^2=\sq_{x|y}-\k_0\sq_{\psi|y}$, $\hat\k=\k_0$, and $I_{\rm acquired}=I(X;Y)$.

(ii) if $\k^2_0\sq_{\psi|y}>\sq_{x|y}$, we have necessarily $D(P_{X|Y}\|\pi\ast G_{\sq_{\psi|y}})>0$, and $\hat\s^2_\pi=0$ but, generally, $\hat\k\neq\k_0$.\\

An illustration of this second case is provided by a model where $y_t=x_t$ but $\psi_t\neq x_t$, i.e., $\sq_{y|x}=0$ and $\sq_{\psi|y}=\sq_{\psi|x}$. For this particular model, the value of acquired information is (see Appendix~\ref{app:sensor}):
\beq\label{eq:solsensor}
I_{\rm acquired}=\frac{1}{2}\left(\hat\k-\ln(1-\hat\k)\right),\quad{\rm with}\quad \hat\k=\frac{\sqrt{\zeta(\zeta+4)}-\zeta}{2}\quad{\rm and}\quad \zeta=\frac{\sq_x}{\sq_{\psi|x}}.
\eeq
This formula shows that the value of acquired information can be strictly larger than the mutual information between the input and output of the sensor $C$, since 
\beq
I_{\rm acquired}\geq I(X;\Psi)=\frac{1}{2}\ln(1+\zeta)
\eeq
with equality if and only if $\zeta=0$ (Figure~\ref{fig:ei}C).

\section{From evolutionary dynamics to thermodynamics}

The problem of formalizing and quantifying the notion of information also lies at the foundations of thermodynamics. As pointed out by Maxwell in a famous thought experiment, an intelligent being may take advantage of microscopic measurements to extract work from a single heat bath, in apparent contradiction with the second law of thermodynamics~\cite{Leff02}. Maxwell's demon is today at the center of an active field of research, stochastic thermodynamics, where many results involve information theoretic quantities~\cite{Parrondo:2015cv}. Recently, Vinkler, Permuter and Merhav showed that the two problems of optimizing the growth rate of a population and optimizing the work extracted from a feedback-controlled thermodynamical system are formally related~\cite{Vinkler:2014vc}. Here, we present and develop this analogy, first with a simple two-state model, then with more generic discrete and Gaussian models.

\subsection{Simple two-state system}

\begin{figure}
\centering
\includegraphics[width=.3\linewidth]{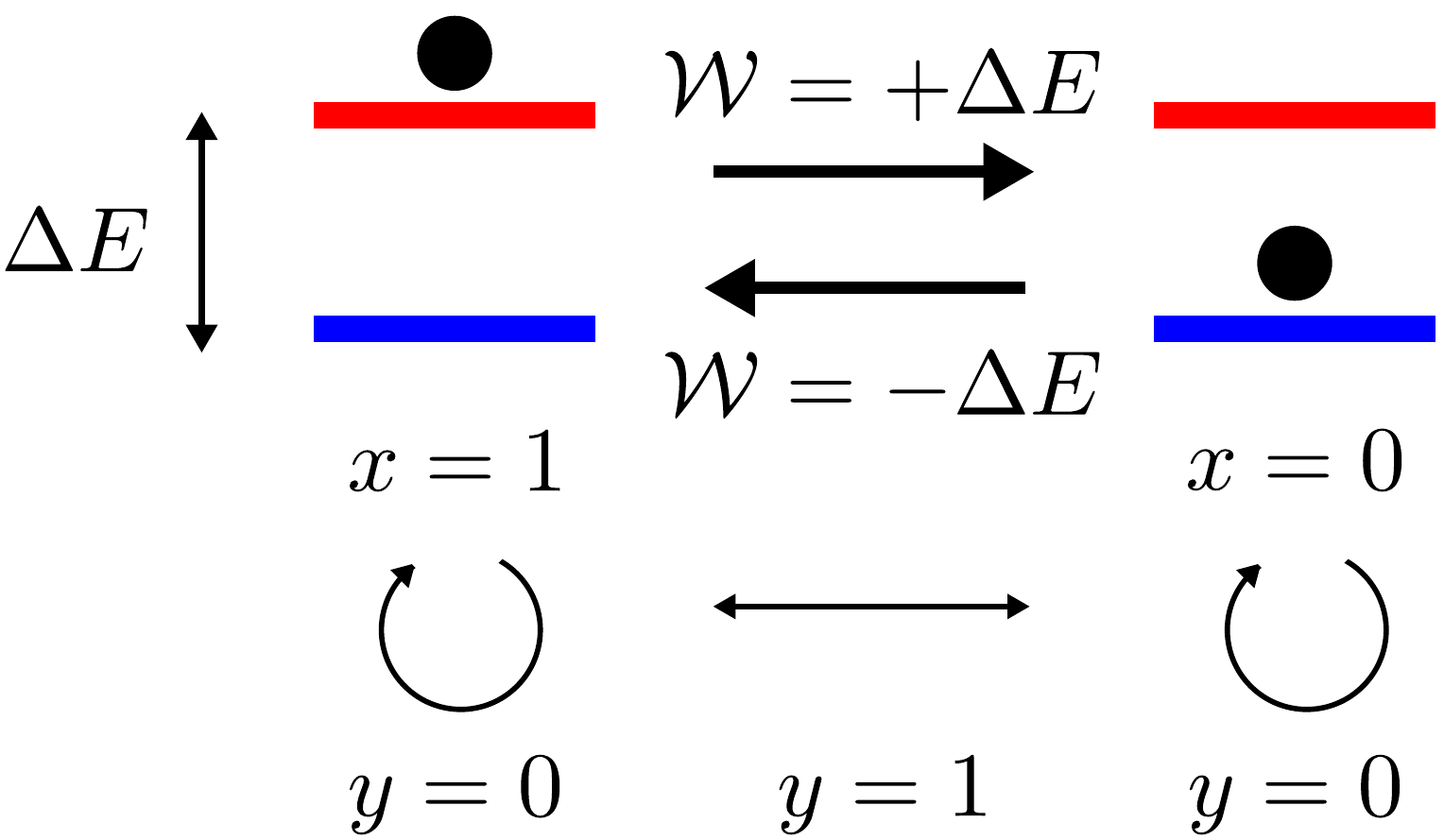}
\caption{Optimal control of a two-level system -- A particle at equilibrium with a heat bath can be in two states: an up state with energy $+\Delta E$ ($x=1$) or a down state with energy $0$ ($x=0$). A measurement is made which indicates, with an error rate $\e<1/2$, whether the particle is up ($y=1$) or down ($y=0$). Based on this measurement, a demon can chose to switch the two levels, thus extracting a work $\W=+\Delta E$ if the particle was up and performing a work $\W=-\Delta E$ if it was down. To extract a maximal work in average, the optimal strategy of the demon is to switch the two levels if and only if the particle is measured in the up state ($y=1$), as indicated in the bottom.\label{fig:demon}}
\end{figure}

As one of the simplest thermodynamical systems with feedback control, we consider a model where a particle can be in two states, either ``down'' in potential $V=0$ or ``up'' in potential $V=\Delta E>0$ (Figure~\ref{fig:demon}). The particle is initially at thermal equilibrium with a heat bath at inverse temperature $\beta=1/(k_BT)$, so that it has probability $p_0=1/(1+e^{-\beta \Delta E})$ to be in the down state, and probability $p_1=1-p_0$ to be in the up state. At regular intervals of time $\tau$, long compared to the equilibration time, a demon can chose to suddenly switch the two levels, thus bringing down the particle if it was up and up if it was down. In doing so, he can extract a work $\W=+\Delta E$ if the particle was in the up state, while losing $\W=-\Delta E$ otherwise. In absence of information on the location of the particle, the expected outcome of the operation is $\E[\W]= (p_1-p_0)\Delta E<0$, a negative result in agreement with the impossibility to extract work from a single heat bath. If the demon knows exactly the location of the particle, on the other hand, he can decide to switch the potential only when the particle is in the up state. As it happens with probability $p_1$, he can thus expect to extract a positive work, $\E[\W]= p_1\Delta E>0$. In the intermediate situation, which we now examine, the demon makes a noisy measurement of the location of the particle and must devise a strategy to optimize the extracted work.\\

To formalize the problem, let denote by $x$ the state of the particle at the time of a measurement, with $x=1$ if it is in the up state and $x=0$ otherwise (Figure~\ref{fig:demon}). Immediately before the demon makes a decision to switch or not the potential, the particle has thus an energy $E_0(x)=x\Delta E$. If $\phi=1$ denotes the choice to switch the potential and $\phi=0$ the choice to leave it unchanged, the energy of the particle after making and implementing choice $\phi$ is $E_1(x|\phi)=|\phi-x|\Delta E$ and the extracted work is $\W(x,\phi)=E_0(x)-E_1(x|\phi)=(2x-1)\phi \Delta E$. Let now consider the outcome $y$ of a measurement of $x$, whose noise is characterized by a conditional probability $P_{Y|X}(y|x)$; for instance, $x= y$ with probability $1-\epsilon$, but $x=1-y$ with an error rate $\epsilon$ (binary symmetric channel). A strategy choosing $\phi$ given $y$ with probability $\rho(\phi|y)$ will extract a mean work
\beq\label{eq:Wb}
\E[\W]=\E_{X,Y}\E_\Phi[\W(X,\Phi)|Y]=\sum_{x,y,\phi}P_X(x)P_{Y|X}(y|x)\rho(\phi|y)\W(\phi,x),
\eeq
where $P_X(x)=e^{-\beta E_0(x)}/Z_0$ is the equilibrium distribution that describes the particle at the time of the measurement, with $Z_0=1+e^{-\beta \Delta E}$. Because of the linearity of Eq.~\eqref{eq:Wb}, the question of finding a strategy $\hat\rho(\phi|y)$ that optimizes the mean extracted work $\E[\W]$ has a trivial answer: it is simply to switch the potential ($\phi=1$) if and only if the state $x=1$ is the most likely given $y$. With a binary symmetric channel with error rate $\epsilon<1/2$, this corresponds to the pure strategy $\phi=y$, i.e., $\hat\rho(\phi|y)=\delta(\phi,y)$, and results in
\beq\label{eq:Wbopt}
\E[\W]=\E_{X,Y}[\W(X;Y)|Y]=\sum_{x,y}P_X(x)P_{Y|X}(y|x)\W(y,x).
\eeq
More generally, the outcome of a measurement determines the optimal decision, $\phi=\phi(y)$, and without loss of generality we can assume that the signal directly indicates the optimal choice, $y=\phi$.\\

An analogy with models of population dynamics arises when introducing the conditional probability $\pi(x|y)= e^{-\beta E_1(x|y)}/Z_1(y)$, where $Z_1(y)=\sum_x e^{-\beta E_1(x|y)}$~\cite{Vinkler:2014vc}. Because $Z_1(y)=\sum_x e^{-\beta E_0(x)}=Z_0$ for all $y$, we can indeed write the extracted work as
\beq\label{eq:W0}
\W(x,y)=E_0(x)-E_1(x|y)=\beta^{-1}\ln\frac{\pi(x|y)}{P_X(x)},
\eeq
and, after averaging,
\beq\label{eq:EW0}
\E[\W]=\beta^{-1}\ \E_{X,Y}\left[\ln\frac{\pi(X|Y)}{P_X(X)}\right].
\eeq
Up to a multiplying factor $\beta$, this expression is formally identical to the expression for the growth rate $\L$ of a discrete Kelly model given in Eq.~\eqref{eq:Lkelly}, with $K(x)=1/P_X(x)$. This particular value of $K(x)$ has a simple interpretation in gambling: it defines a fair game, with $\hat\L=\E_X\ln K(X)-H(X)=0$. From the standpoint of Kelly's model, the choice of a potential $E_1(x|y)$ thus appears as the choice of a strategy. Following Eq.~\eqref{eq:Lhid}, the mean extracted work satisfies
\beq\label{eq:EW}
\beta\E[\W] =I(X;Y)-\E_Y[D(P_{X|Y}(.|Y)\| \pi(.|Y))].
\eeq
Irrespectively of the measurement scheme, the extracted work is therefore bounded by the mutual information between the actual and measured locations of the particle: $\beta\E[\W] \leq I(X;Y)$.

\subsection{General discrete systems}

Reaching the bound $\beta\E[\W] =I(X;Y)$ requires a potential $E_1(x|y)$ verifying $e^{-\beta E_1(x|y)}/Z_1(y)=P_{X|Y}(x|y)$. This potential, however, need not satisfy $Z_1(y)=Z_0$ for all $y$. Introducing the free energies $F_0=-\beta^{-1}\ln Z_0$ and $F_1(y)=-\beta^{-1}\ln Z_1(y)$, the expression for the extracted work when the particle is in $x$ and the measurement indicates $y$, Eq.~\eqref{eq:W0}, generalizes to
\beq
\W_0(x,y)=E_0(x)-E_1(x|y)=\beta^{-1}\ \ln\frac{\pi(x|y)}{P_X(x)}+F_0-F_1(y).
\eeq
In average, the demon will thus extract
\beq\label{eq:2ndlawfeedback}
\E[\W_0]=\beta^{-1}\left(I(X;Y)-\E_Y[D(P_{X|Y}(.|Y)\| \pi(.|Y))]\right)-\Delta F,
\eeq
where $\Delta F=\E_Y[F_1(Y)]-F_0$. This quantity is analogous to a difference of free energies, but note that the state of the system immediately after the operation is generally not be an equilibrium state. Eq.~\eqref{eq:2ndlawfeedback} implies the inequality
\beq
\E[\W_0]\leq\beta^{-1}I(X;Y)-\Delta F.
\eeq
This inequality corresponds to a known generalization of the second law of thermodynamics in presence of feedback~\cite{Sagawa:2008ev}. It is more frequently written $\E[\W_0^-]-\Delta F\geq-\beta^{-1} I(X;Y)$, where $\W_0^-=-\W_0$ is the average work {\it performed} on the system~\cite{Parrondo:2015cv}. We follow here the opposite convention of counting positively the extracted work for consistency with the sign of the growth rate in the evolutionary model.\\

To define a cyclic process, the particle needs to be brought back to equilibrium in $E_0(x)$. To this end, the demon has to perform a work $\W_1^-(y)\geq F_0-F_1(y)+\W_{\rm irr}^-(y)$ where the irreversible work $\W_{\rm irr}^-(y)=-\beta^{-1}D(P_{X|Y}(.|y)\|\pi(.|y))$ is non-zero when the distribution $P_{X|Y}(x|y)$ of the particle immediately after the measurement differs from the equilibrium distribution in the potential $E_1(x|y)$~\cite{Esposito:2011he}. This work $\W_1^-(y)$ performed on the system is to be subtracted from the extracted work $\W_0(x,y)$ when estimating the net extracted work over a complete cycle, $\W(x,y)=\W_0(x,y)-\W_1^-(y)$. In average, this results in an extracted work satisfying
\beq
\beta\E[\W] \leq I(X;Y).
\eeq
This inequality becomes an equality if $\pi=P_{X|Y}$ and the restoration of the original potential is quasi-static, a protocol known to be optimal for discrete-feedback thermodynamic engines~\cite{Horowitz:2011ij}.\\

The mapping presented so far is to Kelly's model, which corresponds to taking two limits in the evolutionary model, a limit of perfect selectivity and a limit of no inheritance. We now examine how the analogy may be extended beyond these two limits.

\subsection{Inheritance}

Extensions to include inheritance (better called ``memory'' in this context) are considered in~\cite{Vinkler:2014vc}. A direct mapping to an evolutionary model with inheritance but perfect selectivity is to assume a multi-step process in which the system is brought back to equilibrium in a new potential $E_0^t(x)$  every time, where $E_0^t(x)$ differs but is correlated to $E_0^{t-1}(x)$. A more interesting extension, however, is to consider that the particle does not equilibrate with the thermal bath before a new measurement and change of potential are made. Physically, equilibration takes time and instead of extracting a maximal work $\E[\W]$, it may be more desirable to extract a maximal power $\mathcal{P}=\E[\W]/\tau$, where $\tau$, the  time taken by a cycle, may itself be optimized. We present in Appendix~\ref{app:feedback} an extension of Eq.~\eqref{eq:2ndlawfeedback} to cover such non-equilibrium protocols. While the extracted work can still be written with information theoretic quantities, their interpretation is complicated by the fact that the state $x_t$ of the system prior to a measurement now depends on the series of choices $\phi^{t-1}=(\phi_1,\dots,\phi_{t-1})$ made by the demon. From the standpoint of the evolutionary model, this corresponds to a feedback from the state of the population to the state of the environment, a biologically relevant phenomenon that could be further studied within the present framework.

\subsection{Finite selectivity}

A mapping to an evolutionary model with finite selectivity is for instance obtained by assuming a separation of scales between a macro-state $x$, which is measured, and micro-states $\phi$, which are manipulated, with $S(\phi,x)$ representing the density of states, i.e., the number of micro-states $\phi$ associated with the macro-state $x$. In this mapping, the demon makes a macroscopic measurement of $x$ but, given the result $y$, can tune every microscopic energy levels from $E_0(\phi)$ to $E_1(\phi|y)$. Assuming that we start and end with the micro-states at equilibrium given their macro-state, Eq.~\eqref{eq:W0} becomes $\W_0(x,y)\leq\F_0(x)-\F_1(x|y)$ where $\F_0(x)=-\beta^{-1}\ln(\sum_\phi S(\phi,x)e^{-\beta E_0(\phi)})$ is the free energy of a system at equilibrium in macro-state $x$, and $\F_1(x|y)=-\beta^{-1}\ln(\sum_\phi S(\phi,x)e^{-\beta E_1(\phi|y)})$ at equilibrium in the new distribution of energy levels. By writing again $e^{-\beta E_1(\phi|y)}=\pi(\phi|y)e^{-\beta F_1(y)}$, we obtain
\beq
\W_0(x,y)\leq \F_0(x)+\beta^{-1}\ln(\sum_\phi S(\phi,x)\pi(\phi|y))-F_1(y),
\eeq
and therefore 
\beq
\E[\W_0]\leq \E_X[\F_0(X)]+\beta^{-1}\E_{X,Y}[\ln(\sum_\phi S(\phi,X)\pi(\phi|Y))]-\E_Y[F_1(Y)],
\eeq
with equality if the energy levels are changed quasi-statically. The conditional probability $\pi$ is involved in the last two terms of the right-hand side, but, as in the two-state model of Figure~\ref{fig:demon}, we may assume that the demon is constrained to $F_1(y)=F_0$ for all $y$ and that the last term is therefore independent of $\pi$. In this case, the problem of choosing the energy levels $E_1(\phi|y)$ so as to optimize the extracted work is formally identical to the problem of optimizing the growth rate of an evolutionary model with finite selectivity.

\subsection{Gaussian systems}

Mapped to its thermodynamical analog, the Gaussian model of evolutionary dynamics becomes the problem of controlling a Brownian particle with harmonic potentials. The Gaussian distribution $P_X(x)=G_{\sq_x}(x)$ is indeed the equilibrium distribution of a particle in contact with an heat bath at inverse temperature $\beta$ and in a potential $V_0(x)=kx^2/2$ when considering $\sq_x=(\beta k)^{-1}$. In the simplest version of the analogy, a demon observes a particle at equilibrium in this potential and measures its location $x$ at $y$, with a noise characterized by $P_{Y|X}(y|x)=G_{\sq_{y|x}}(y-x)$. His problem is then to change the potential to $V_1(x|y)$ so as to extract a maximal work.\\

While changing the stiffness $k$ of the potential may allow the demon to extract more work, the simplest scenario is when only translations are allowed, from $V_0(x)=kx^2/2$ to $V_1(x|y)=k(x-\phi_1)^2/2$, a case where $Z_1(y)=Z_0=(2\pi\sq_x)^{1/2}$, and therefore $\Delta F=0$ in Eq.~\eqref{eq:2ndlawfeedback}. As a consequence of the formal mapping to an evolutionary model, the optimal strategy of the demon is to move the potential to $\phi_1=\hat\k y_0$ with $\hat\k$ given by Eq.~\eqref{eq:Kgain}, i.e., $\hat\k=1/(1+\sq_{y|x}/\sq_x)$. The optimal extracted work is the value of acquired information given by Eq.~\eqref{eq:Iacqfs} when taking $\sq_s=\sq_x$: $\E[\hat W]=(1-\sq_{x|y}/\sq_x)/2=(1+\sq_{y|x}/\sq_x)^{-1}/2$. These expressions corresponds to those obtained by a more direct calculation~\cite{Abreu:2011ft}.\\ 

If the process is repeated after quasi-statically restoring the potential at a location that is correlated but differs from its original location, the problem maps to the Gaussian model of evolutionary dynamics with inheritance. Specifically, it corresponds to beginning each cycle $t$ with the particle at equilibrium in $V_t(x)=k(x-x_t)^2/2$, where $x_t=ax_{t-1}+\nu_t$ and where $\nu_t$ is normally distributed with variance $\sq_{x_1|x_0}$. This problem also maps to a problem of stochastic control solved by Kalman~\cite{Kalman:1960tn}. In Kalman's model, the state $x_t$ of a system, its measured state $y_t$ and its estimated state $\phi_t$ are assumed to follow the recursions
\bea
x_t&=&ax_{t-1}+\nu_t,\qquad \nu_t \sim \mathcal{N}(0,\sigma_{x_1|x_0}^2),\\
y_t&=&x_t+\nu'_t,\qquad\quad\ \  \nu'_t \sim \mathcal{N}(0,\sigma_{y_1|x_1}^2),\\
\phi_t &=&\lambda \phi_{t-1}+\kappa y_t,
\eea
and the objective is to find the estimation $\phi_t$ that minimizes the mean square error $\E[(\phi_t-x_t)^2]$ by choosing appropriately the two parameters $\l$ and $\k$. A standard application is for instance to tracking, where the current position and velocity of a target must be estimated from past estimations and from independent measurements. The optimal values for $\hat\l$ and $\hat k$ are also given by Eq.~\eqref{eq:Kgain} (as for our model, a generalization  to multidimensional variables is straightforward).\\

A physically more interesting situation is when the particle has no time to equilibrate before a new measurement and manipulation are made. The Gaussian setting is here again well-suited for making explicit calculations of the maximal work that may be extracted with such non-equilibrium protocols~\cite{Bauer:2012ks} (see also Appendix~\ref{app:noneq}). The results obtained for Brownian particles in harmonic potentials suggest that the feedback of a population onto its environment could also be studied analytically in Gaussian models of population dynamics.

\section{Discussion}

We reviewed an approach to quantify the value of informations in evolution by analyzing abstract models of population dynamics, and showed how analytical expressions can be obtained when considering a particular Gaussian limit. This approach illustrates how the value of an information may depend on factors beyond the characteristics of the channel that directly conveys it. In particular, it shows how the value of an information acquired from the current environment is tied to the value of the information inherited from previous generations. Alternative approaches for quantifying information are possible, for instance based on well-chosen sets of axioms~\cite{Csiszar:2008wd}, but at the risk of omitting an important feature of the problem. Although elementary, our model indicates that several constraints should generically be taken into account, including causality, selectivity of the environment and individual stochasticity. Studies of informations in thermodynamics take a similar approach of analyzing simple models and also find that different quantities for quantifying information may arise depending on the protocol~\cite{Horowitz:2014ev}. Remarkably, a similar mathematical formalism emerges from the two problems~\cite{Vinkler:2014vc}.\\

This formal correspondence suggests that methods and concepts may be transferred between disciplines. In~\cite{Vinkler:2014vc}, the authors thus applied the concept of universal strategy from information theory~\cite{Cover:1996we} to devise a thermodynamical protocol that optimally extracts work when the statistical properties of the system, for instance the characteristics of the information channel, are unknown. Reciprocally, many results have been obtained recently in stochastic thermodynamics~\cite{Parrondo:2015cv} which may provide new insights on evolutionary dynamics. For instance, inequalities on  the mean extracted work are known to generalize to fluctuation theorems, which take into account fluctuations around the mean result and connect macroscopic observations to the underlying time-reversal symmetry of the microscopic dynamics. Given the analogy between extracted work and growth rate, similar relations may hold for population dynamics. One such fluctuation relation has in fact already been established for evolutionary dynamics by Mustonen and L\"assig~\cite{Mustonen:2010ig}, but at a different level of analysis: they considered fluctuations arising from finite population sizes, which are ignored in the present analysis of our models. The path integral formalism at the core of their approach has, however, its counterpart at our level of analysis~\cite{Kussell:2005tk}.\\

Another challenge is to move beyond the formal analogy towards an integrated treatment of evolutionary and thermodynamical constraints. The presented models account for part of the evolutionary constraints but the information processor $\pi$, the sensor $C$ and the ``replicator'' $S$ are introduced as ad-hoc parameters, with no reference to physics or evolution. Several recent studies have investigated thermodynamical constraints on information processing~\cite{Lan:2012in}, biochemical sensing~\cite{Lang:2014ir} or replication~\cite{England:2013ed}, and others have investigated evolutionary constraints at the inter-molecular~\cite{Francois:2014hg} and intra-molecular~\cite{Hemery:2015ei} levels. Given the interplay between local and global properties that simple models already exhibit, integrating these different constraints appears as both necessary and interesting.

\begin{acknowledgments}
I thank B. Houchmandzadeh and M. Ribezzi for helpful comments.
\end{acknowledgments}

\section*{APPENDICES}

\appendix

\section{Mapping from Eq.~\eqref{eq:gp} to Eq.~\eqref{eq:basic}}\label{app:mapping}

The model described by Eq.~\eqref{eq:gp} is mapped to the model described by Eq.~\eqref{eq:basic} by defining
\beq
\tilde\phi_t=(\g_t,\phi_t),\quad \tilde x_t=(0,x_t),\quad \tilde y_t=(y_t,z_t)
\eeq
and
\beq
\pi(\tilde\phi_t|\tilde\phi_{t-1},\tilde y_t)=H(\tilde\phi^1_t|\tilde\phi^1_{t-1},\tilde\phi^2_t,\tilde y^2_t)D(\tilde\phi^2_t|\tilde\phi^1_{t-1},\tilde y^1_t),\qquad \tilde S(\tilde\phi_t,\tilde x_t)=S(\tilde\phi^2_t,\tilde x^2_t),
\eeq
where $\tilde\phi^k_t$ corresponds to the $k$-th component of $\tilde\phi_t$, i.e., $\tilde\phi_t=(\tilde\phi_t^1,\tilde\phi_t^2)$. 
Note that $\tilde S$ is of the form $\tilde S=\tilde K\tilde \Delta$ as in Eq.~\eqref{eq:S=KD} if $S$ is itself of the form $S=K\Delta$.

\section{Decomposition of the growth rate}\label{app:decamp}

We detail here the decomposition of the growth rate given in Eq.~\eqref{eq:hatLtildePi} for the discrete model in absence of inheritance. The idea is to write
\beq
\begin{split}
\L&=\E_{X,Y}[\ln(K(X)\tilde\pi(X|Y))]\\
&=\E_{X,Y}[\ln(K(X)]+\E_{X,Y}[\ln P_X(x)]+\E_{X,Y}\left[\ln \frac{P_{X|Y}(X|Y)}{P_X(X)}\right]+\E_{X,Y}\left[\ln\frac{\tilde\pi(X|Y)}{P_{X|Y}(X|Y)}\right],
\end{split}
\eeq
and to recognize that $\E_{X,Y}[\ln P_X(x)]=-H(X)$, $\E_{X,Y}[\ln P_{X|Y}(X|Y)/P_X(X)]=I(X;Y)$ and 
\beq
\begin{split}
\E_{X,Y}\left[\ln\frac{\tilde\pi(X|Y)}{P_{X|Y}(X|Y)}\right]&=\sum_{x,y}P_{X,Y}(x,y)\ln\frac{\tilde\pi(x|y)}{P_{X|Y}(x|y)}=\sum_y P_Y(y)\sum_{x}P_{X|Y}(x|y)\ln\frac{\tilde\pi(x|y)}{P_{X|Y}(x|y)}\\
&=-\sum_y P_Y(y)D(P_{X|Y}(.|y)\| \tilde\pi(.|y))=-\E_Y[D(P_{X|Y}(.|Y)\| \tilde\pi(.|Y))].
\end{split}
\eeq

\section{Gaussian random variables}\label{app:general}

A Gaussian random variable $X$ is characterized by its mean $x_0=\E[X]$ and its variance $\sq_x=\E[X^2]-\E[X]^2$, and its probability density is $P_X(x)=G_{\sq_x}(x-x_0)$, where $G_{\sq_x}(x)=(2\pi\sq_x)^{-1/2}\exp(-x^2/2\sq_x)$.\\

Its differential entropy $h(X)=-\int\ud xP_X(x)\ln P_X(x)$ is
\beq
h(X)=\frac{1}{2}\ln(2\pi e\sq_x).
\eeq

The mutual information $I(X;Y)=h(X)-h(X|Y)$ between $X$ and another Gaussian random variable $Y$ whose conditional probability given $x$ is $P_{Y|X}(y|x)=G_{\sq_{y|x}}(y-x)$ is
\beq
I(X;Y)=\frac{1}{2}\ln\left(1+\frac{\sq_x}{\sq_{y|x}}\right).
\eeq\\

The relative entropy between two Gaussian probability densities is
\beq\label{eq:DKLG}
D(G_{\sq_0}(.-x_0)\|G_{\sq_1}(.-x_1))=\frac{1}{2}\left(\frac{\sq_0+(x_1-x_0)^2}{\sq_1}-\ln\frac{\sq_0}{\sq_1}-1\right).
\eeq\\

Finally, given $P_{X_1|X_0}(x_1|x_0)=G_{\sq_{x_1|x_0}}(x_1-ax_0)$ and $P_{Y_1|X_1}(y_1|x_1)=G_{\sq_{y_1|x_1}}(y_1-x_1)$, the conditional probability $P_{X_1|Y_1,X_0}$, which by Bayes' rule is proportional to $P_{Y_1|X_1}P_{X_1|X_0}$, is also Gaussian and given by
\beq\label{eq:BG}
P_{X_1|Y_1,X_0}(x_1|y_1,x_0)=G_{\sq_{x_1|y_1,x_0}}(x_1-\l x_0-\k y_1),
\eeq
with
\beq
\k=\frac{1}{1+\sq_{y_1|x_1}/\sq_{x_1|x_0}},\qquad \l=a(1-\k),\qquad \sq_{x_1|y_1,x_0}=\k\sq_{x_1|x_0},
\eeq
or, equivalently, $\s^{-2}_{x_1|y_1,x_0}=\s_{x_1|x_0}^{-2}+\s_{y_1|x_1}^{-2}$.

\section{Growth rate of the Gaussian model}\label{app:LG}

Eq.~\eqref{eq:L} for the growth rate $\L$ of the Gaussian model is obtained by considering 
\beq
n_t(\phi_t)=\frac{1}{W_t}K(x_t)\int\ud\phi_{t-1}\ G_{\sq_s}(\phi_t-x_t)G_{\sq_\pi}(\phi_t-\l\phi_{t-1}-\k y_t)n_t(\phi_{t-1}),
\eeq
with $n_t(\phi_t)$ of the form $n_t(\phi_t)=G_{\sq_t}(\phi_t-m_t)$, which leads to
\bea
m_t&=&\frac{\sq_s}{\sq_s+\sq_\pi+\l^2\sq_{t-1}}(\l m_{t-1}+\k y_t)+\frac{\sq_\pi+\l^2\sq_{t-1}}{\sq_s+\sq_\pi+\l^2\sq_{t-1}}x_t\\
\sq_t&=&(\s_s^{-2}+(\sq_\pi+\l^2\sq_{t-1})^{-1})^{-1}\\
W_t&=&K(x_t)G_{\sq_s+\sq_\pi+\l^2\sq_{t-1}}(\l m_{t-1}-x_t+\k y_t).
\eea
The variance $\sq_t$ has a fixed point $\sq_\infty$ in terms of which the growth rate can be rewritten as
\beq\label{eq:LLL}
\L=\lim_{t\to\infty}\E[\ln W_t]=\L^*-\frac{1}{2}\ln(2\pi\sq_s)+\frac{1}{2}\ln\frac{\a}{\l}-\frac{\a}{2\l\sq_s}\lim_{t\to\infty}\E[z^2_t],
\eeq
where $\L^*=\E_X[\ln K(X)]$, 
\beq
z_t=\l m_{t-1}-x_t+\k y_t,
\eeq
and
\beq
\alpha=\frac{\l\sq_s}{\sq_s+\sq_\pi+\l^2\sq_\infty}=\frac{2\l}{1+\l^2+\b+((1-\l^2-\b)^2+4\b)^{1/2}},,\quad{\rm with}\quad \b=\frac{\sq_\pi}{\sq_s}.
\eeq
Given that $x_{t+1}=ax_t+b_t$ and $y_{t+1}=x_{t+1}+b'_{t+1}$ with $b_t\sim\N(0,\sq_{x_1|x_0})$ and $b'_{t+1}\sim\N(0,\sq_{y_1|x_1})$, we have
\beq
z_{t+1}=\alpha z_t+\epsilon x_t+(\k-1)b_t+\k b'_{t+1},\quad{\rm with}\quad\e= \l-a(1-\k).
\eeq
Using $\sum_{k=0}^t\a^{t-k}x_k=\sum_{k=0}^t(\a^{t-k}-a^{t-k})/(\a-a)b_k$, we obtain
\beq
z_{t+1}=\frac{1}{\a-a}\sum_{k=0}^t(\d \a^{t-k}-\e a^{t-k})b_k+\k\sum_{k=0}^t\a^{t-k}b'_{k+1},\quad{\rm with}\quad \d= \l-\a(1-\k),
\eeq
and, since the $b_k$ and $b'_k$ are all independent, with variances $\E[b_k^2]=\sq_{x_1|x_0}$ and $\E[b_k'^2]=\sq_{y_1|x_1}$,
\bea
\lim_{t\to\infty}\E[z^2_{t+1}]&=&\frac{1}{(\a-a)^2}\left(\frac{\d^2}{1-\a^2}-\frac{2\d\e}{1-a\a}+\frac{\e^2}{1-a^2}\right)\sq_{x_1|x_0}+\k^2\frac{\sq_{y_1|x_1}}{1-\a^2}\\
&=&\frac{(\l^2+(1-\k)^2)(1+a\a)-2\l(1-\k)(a+\a)}{(1-\a^2)(1-a\a)(1-a^2)}\sq_{x_1|x_0}+\k^2\frac{\sq_{y_1|x_1}}{1-\a^2}.
\eea
Plugged into Eq.~\eqref{eq:LLL}, it leads to Eq.~\eqref{eq:L}.

\section{Decomposition of the growth rate of the Gaussian model}\label{app:2form}

Since the Gaussian model can be obtained as a continuous limit of the discrete model, Eqs.~\eqref{eq:Lnoinh}-\eqref{eq:Ginfsel} directly result from Eqs.~\eqref{eq:hatLtildePi}-\eqref{eq:LKinheritance} by taking the same limit. The decomposition can also be derived directly from the general formula of Eq.~\eqref{eq:L} as we illustrate it here in the simplest case where the two limits are taken.\\

The first limit, of perfect selectivity, corresponds to $\sq_s\to 0$, such that Eq.~\eqref{eq:L} becomes
\beq\label{eq:GK}
\L=\L^*-\frac{1}{2}\ln(2\pi\sq_\pi)-\frac{1}{2\sq_\pi}\left[\frac{\l^2+(1-\k)^2-2\l(1-\k)a}{1-a^2}\sq_{x_1|x_0}+\k^2\sq_{y_1|x_1}\right].
\eeq
The second limit, of no inheritance, simply corresponds to setting $\l= 0$ in this equation, so that
\beq\label{eq:GKelly}
\L=\L^*-\frac{1}{2}\ln(2\pi\sq_\pi)-\frac{1}{2\sq_\pi}\left[(1-\k)^2\sq_{x_1}+\k^2\sq_{y_1|x_1}\right],
\eeq
where $\sq_{x_1}=\sq_{x_1|x_0}/(1-a^2)$ represents the stationary variance of the environmental process, $\sq_{x_1}=\E[X_1^2]$. The optimal strategy $\hat\pi$ is obtained by optimizing $\L$ over $\k$ and $\sq_\pi$, which leads to
\beq
\hat\k=\frac{1}{1+\sq_{y_1|x_1}/\sq_{x_1}},\qquad \hat\s_\pi^2=\hat\k\sq_{y_1|x_1}.
\eeq
$\hat\kappa$ can also be written $\hat\k=\sq_{x_1}/\sq_{y_1}$ where $\sq_{y_1}=\sq_{x_1}+\sq_{y_1|x_1}$ represents the stationary variance of $y_t$.  As expected from the analysis of the discrete model, we verify that the optimal strategy implements a Bayesian estimation, i.e., $\hat\pi=P_{X|Y}$ [see Appendix~\ref{app:general}]. We also verify that the optimal optimal growth rate,
\beq
\hat \L=\L^*-\frac{1}{2}\ln\left(2\pi\frac{\sq_{y_1|x_1}\sq_{x_1}}{\sq_{x_1}+\sq_{y_1|x_1}}\right)-\frac{1}{2},
\eeq
is equivalently written
\beq\label{eq:hatLgauss}
\hat \L=\L^*-h(X)+I(X;Y),
\eeq
where
\beq
h(X)=\frac{1}{2}\ln(2\pi e\sq_{x_1}),\qquad {\rm and}\qquad  I(X;Y)=\frac{1}{2}\ln\left(1+\frac{\sq_{x_1}}{\sq_{y_1|x_1}}\right).
\eeq

More generally, by introducing $\s_{x_1|y_1}^{-2}=\s_{y_1|x_1}^{-2}+\s_{x_1}^{-2}$, so that $(1-\k)^2\sq_{x_1}+\k^2\sq_{y_1|x_1}=\sq_{x_1|y_1}+(\hat\k-\k)^2\sq_{y_1}$, we verify that Eq.~\eqref{eq:GKelly} is equivalent to
\beq
\L=\hat\L-\frac{1}{2}\left[\frac{\sq_{x_1|y_1}+(\hat\k-\k)^2\sq_{y_1}}{\sq_\pi}-\ln\frac{\sq_{x_1|y_1}}{\sq_\pi}-1\right]=\hat\L-\E_Y[D(P_{X|Y}(.|Y)\| \pi(.|Y))],
\eeq
as expected from Eq.~\eqref{eq:Lkelly}.

\section{Gaussian model with individual sensors}\label{app:sensor}

Using the formulae of Appendix~\ref{app:general}, the term to maximize in Eq.~\eqref{eq:IacqsensorG} can be written
\beq
I(X;Y)-\E_Y[D(P_{X|Y}\|G_{\sq_\pi+\k^2 \sq_{\psi|y}}(.-\k Y))]=\frac{1}{2}\ln\left(1+\frac{\sq_x}{\sq_{y|x}}\right)-\frac{1}{2}\left(\frac{\sq_{x|y}+(\k-\k_0)^2\sq_y}{\sq_\pi+\k^2\sq_{\psi|y}}-\ln\frac{\sq_{x|y}}{\sq_\pi+\k^2\sq_{\psi|y}}-1\right).
\eeq
When $Y\to X$, $\k_0\to1$, $\sq_y\to0$, $\sq_{y|x}\to 0$ and $\sq_{x|y}\to 0$ but $\sq_{x|y}/\sq_{y|x}\to 1$ and it simplifies to
\beq
\frac{1}{2}\left[\ln\left(\frac{\sq_x}{\sq_\pi+\k^2\sq_{\psi|x}}\right)-\frac{(\k-1)^2\sq_x}{\sq_\pi+\k^2\sq_{\psi|x}}+1\right].
\eeq
The maximum over $\sq_\pi$ is reached for $\sq_\pi=0$ and taking the derivative with respect to $\k$ leads to
\beq
\k^2\sq_{\psi|x}+(\k-1)\sq_x=0
\eeq
whose solution is given in Eq.~\eqref{eq:solsensor}.

\section{The Gaussian model as a limit of the general model of Ref.~\cite{Rivoire:2014kta}}\label{app:pg}

A general model is defined by Eqs.~[S1]-[S2] in the Supporting Information of~\cite{Rivoire:2014kta}, which we repeat here with only slightly modified notations:
\bea
\g_0'&=&\l_0\g_0+\k_0z_t+\omega_0\phi_0+\nu_H,\quad \nu_H\sim\N(0,\sq_H),\\
\phi_0&=&\theta_0\g_0+\rho_0y_t+\nu_D,\quad\nu_D\sim\N(0,\sq_D),\\
S(\phi_0,x_t)&=&\exp[r_{\rm max}-(\phi_0-x_t)^2/(2\sq_s)],\\
x_t&=&ax_{t-1}+b_t,\quad b_t\sim\N(0,\sq_{x_1|x_0}),\\
y_t&=&x_t+b'_t,\quad b'_t\sim\N(0,\sq_{y|x}),\\
z_t&=&x_t+b''_t,\quad b''_t\sim\N(0,\sq_{z|x}).\\
\eea
Without loss of generality it can be assumed that $\sq_s=1$. The formula for the growth rate of this general model is given with an error in Eq.~[S3] of~\cite{Rivoire:2014kta}. The correct formula is
\beq
\L=r_{\rm max}+\frac{1}{2}\ln\frac{\a}{\eta}-\frac{\a}{2\eta(1-\a^2)}\left[\frac{(\u^2+(1-\rho_0)^2)(1+a\alpha)-2\u(1-\rho_0)(a+\alpha)}{1-a\alpha}\sq_x+\rho_0^2(1-2\a\l_0+\l_0^2)\sq_{y|x}+\k_0^2\theta_0^2\sq_{z|x}\right],
\eeq
where $\sq_x\equiv \sq_{x_1|x_0}/(1-a^2)$, where $\eta$ and $\u$ given by 
\beq
\eta=\l_0(1+\sq_D)+\w_0\theta_0,\qquad\u= (\w_0+\k_0)\theta_0+(1-\rho_0)\l_0,
\eeq
and
\beq
\a=\frac{2\tilde\lambda}{1+\tilde\lambda^2+\tilde\sigma_H^2+\left((1-\tilde\lambda^2-\tilde\sigma_H^2)^2+4\tilde\sigma_H^2\right)^{1/2}},
\eeq
with
\beq
\tilde\sigma_H^2=\left(\sq_H+\frac{\omega_0^2\sq_D}{\sq_D+1}\right)\frac{\theta_0^{2}}{\sq_D+1},\quad
\quad\tilde\l=\l_0+\frac{\theta_0\omega_0}{\sq_D+1}
\eeq
These formulae reduce to Eq.~\eqref{eq:L} when taking $\theta_0=\l$, $\rho_0=\k$, $\w_0=1$, $\sq_D=\sq_\pi$, $\l_0=\k_0=\sq_H=\sq_{z|x}=0$ and $r_{\rm max}=\ln K-(1/2)\ln(2\pi\sq_s)$.

\section{Feedback control out of equilibrium}\label{app:feedback}\label{app:noneq}

The state $x_t$ of a system in contact with a heat bath is measured as $y_t$ at regular intervals of time $\tau$, upon which the potential in which the system evolves is changed from $V_{t-1}(x)$ to $V_t(x)$. This change is done without knowing the current state $x_t$, but may depend on the history of past measurements $y^t=(y_1,\dots,y_t)$ as well as on the history of past states at the time of these measurements, $x^{t-1}=(x_1,\dots,x_{t-1})$. If we assume that the potential is controllable by one or several parameters $\ell$, we therefore consider, in the more general case, that $\ell_t=\ell(y^t,x^{t-1})$ [in more constrained cases, $\ell_t$ may depend only on some of variables, e.g., $\ell_t=\ell(y^t)$ when only the present and past measurements are available]. In-between two measurements, the system relaxes in a constant potential $V_t(x)$ but may not reach equilibrium; its dynamics is generally stochastic, due to the interaction with the heat bath, and may for instance be described by a Master equation with rates satisfying detailed balance. When changing the potential from $V_{t-1}(x)$ to $V_t(x)$, a demon extracts a work $\W_t=V_{t-1}(x_t)-V_{t}(x_t)$. The goal of the demon is either to optimize the total extracted work $\W_{\rm tot}=\E[\sum_t\W_t]$ or, if $\tau$ itself is controllable, to optimize the power $\W_{\rm tot}/\tau$.\\

To formalize this problem, we denote by $p^{\tau}_{t-1}(x_t)$ the probability of the system to be in state $x_t$ at the time of the $t$-th measurement: this probability depends explicitly only on $\ell_{t-1}$ and $x_{t-1}$, which characterize, respectively, the potential $V_{t-1}(x)$ and the state of the system when this potential is switched on. Introducing $F_t=-\beta^{-1}\ln\sum_x e^{-\beta V_t(x)}$ and $p^{(\infty)}_t(x)=e^{\beta [F_t-V_t(x)]}$ (also denoted $\pi$ in the main text), the extracted work may be decomposed as
\beq
 \W_t(x^t,y^t)=V_{t-1}(x_t)-V_t(x_t)=\beta^{-1}\ln\frac{p^{\infty}_t(x_t)}{p^{\tau}_{t-1}(x_t)}+\beta^{-1}\ln\frac{p^{\tau}_{t-1}(x_t)}{p^{\infty}_{t-1}(x_t)}-(F_t-F_{t-1}).
\eeq
We now consider the past history $(x^{t-1},y^{t-1})$ as given and average over $(X_t,Y_t)$ to define
\beq
\E_t[\W_t]= \E_{X_t,Y_t|X^{t-1}=x^{t-1},Y^{t-1}=y^{t-1}}[\W_t(X^t,Y^t)].
\eeq
Since $P_{X_t|X^{t-1},Y^{t-1}}(x_t|x^{t-1},y^{t-1})=p^{\tau}_{t-1}(x_t)$, we have
\beq\label{eq:bEWt}
\beta \E_t[\W_t]=I(X_t;Y_t|x^{t-1},y^{t-1}) -\E [D(q^{\tau}_{t-1}\| p^{\infty}_t)]+\E [D(p^{\tau}_{t-1}\| p^{\infty}_{t-1})]-\beta\E[F_t-F_{t-1}],
\eeq
where
\beq
q_{t-1}^\tau(x_t|y_t)=P_{X_t|X^{t-1},Y^t}(x_t|x^{t-1},y^t)=\frac{P_{Y|X}(y_t|x_t)p_{t-1}^\tau(x_t)}{\sum_x P_{Y|X}(y_t|x)p_{t-1}^\tau(x)}
\eeq
and
\beq
I(X_t;Y_t|x^{t-1},y^{t-1}) =\sum_{x_t,y_t}P_{Y|X}(y_t|x_t)p^{\tau}_{t-1}(x_t)\ln \frac{q_{t-1}^\tau(x_t|y_t)}{p_{t-1}^\tau(x_t)}
\eeq
The total work $\W_{\rm tot}$ is obtained as $\W_{\rm tot}=\sum_t\E_{X^{t-1},Y^{t-1}}[\E_t[\W_t]]$. When $\tau\to\infty$, the third term on the right-hand side of Eq.~\eqref{eq:bEWt} vanishes and we recover the equilibrium result, Eq.~\eqref{eq:2ndlawfeedback}. \\

This formalism can be applied to a Brownian particle in a controllable harmonic potential. For simplicity, we assume that only the location of the potential can be controlled, and its stiffness $k$ is fixed to $k=1$. We also set $\beta=1$. The potential $V_t(x)=(x-\ell_t)^2/2$ is characterized by the location $\ell_t$ of its minimum, and $F_t=F_{t-1}$ for all $t$. We take the relaxation dynamics between measurements to be described by a Fokker-Planck equation,
\beq
\partial_\tau p_t^\tau(x)=\partial_x(\partial_xV_t(x)p_t^\tau(x))+\partial^2_xp_t^\tau(x),
\eeq
with the initial condition is $p_t^0(x)=\delta(x-x_t)$. This equation is easily solved as its solution is Gaussian at all time: $p^\tau_t(x)=G_{\vq_\tau}(x-\mu_t^\tau)$ with
\bea
&\frac{1}{2}\partial_\tau\mu_t^\tau+\mu_t^\tau=\ell_t,\qquad &\mu^0_t=x_t,\\
&\frac{1}{2}\partial_\tau\vq_\tau+\vq_\tau=1,\qquad &\vq_0=0,
\eea
so that
\bea
\mu_t^\tau&=&(1-e^{-\tau})\ell_t+e^{-\tau}x_t,\\
\vq_\tau&=&1-e^{-2\tau}.
\eea
When $\tau\to\infty$, $p^\tau_t(x)$ converges to the equilibrium distribution $p_t^\infty(x)=G_{1}(x-\ell_t)$. Using $P_{Y|X}(y_t|x_t)=G_{\sq_{y|x}}(y_t-x_t)$ and applying Eq.~\eqref{eq:BG}, $q_t^\tau$ is found to be
\beq
q_{t-1}^\tau(x_t|y_t)=G_{\sq_{x|y}}(x_t-(1-\k)\mu^\tau_{t-1}-\k y_t),\qquad \textrm{with}\quad \k=\frac{1}{1+\sq_{y|x}/\vq_\tau},\qquad\sq_{x|y}=\k\sq_{y|x}.
\eeq
The first term in Eq.~\eqref{eq:bEWt} is therefore
\beq
I(X_t;Y_t|x^{t-1},y^{t-1})=\frac{1}{2}\ln\left(1+\frac{\vq_\tau}{\sq_{y|x}}\right).
\eeq
The second term is
\beq\label{eq:Ezt}
\E [D(q^{\tau}_{t-1}\| p^{\infty}_t)]=\frac{1}{2}\left(\sq_{x|y}+\E[z_t^2]-\ln\sq_{x|y}-1\right),
\eeq
with
\beq
z_t =(1-\k)\mu^\tau_{t-1}+\k y_t-\ell_t.
\eeq
The third term is
\beq
\E [D(p^{\tau}_{t-1}\| p^{\infty}_{t-1})]=\frac{1}{2}\left(\vq_\tau+\E[z_t'^2]-\ln\vq_\tau-1\right),
\eeq
with
\beq
z_t'=\mu^\tau_{t-1}-\ell_{t-1}=e^{-\tau}(x_{t-1}-\ell_{t-1}).
\eeq
Given $(x^{t-1},y^{t-1})$, the only term depending on $y_t$ is $\E[z_t^2]$ in Eq.~\eqref{eq:Ezt}. It is optimized by choosing $\ell_t$ so as to have $z_t=0$:
\beq
\hat\ell_t=\k y_t+(1-\k)\mu^\tau_{t-1}=\k y_t+(1-\k)[(1-e^{-\tau})\ell_{t-1}+e^{-\tau}x_{t-1}].
\eeq
By taking $\ell_{t-1}=\hat\ell_{t-1}$, this defines recursively a series of optimal translations $\hat\ell^t$.\\

To express the optimal work, it remains to evaluate $\E[z_t'^2]$ for $\ell^t=\hat\ell^t$.
Since $x_t-\hat\ell_t=(1-\k)(x_t-\mu_{t-1}^\tau)-\k(y_t-x_t)$ where $x_t-\mu_{t-1}^\tau$ and $y_t-x_t$ are statistically independent, we have
\beq
\E[(x_t-\hat\ell_t)^2]=(1-\k)^2\vq_\tau+\k^2\sq_{y|x}=\sq_{x|y},
\eeq
and therefore $\E[z_t'^2]=e^{-2\tau}\sq_{x|y}$. All together, we obtain
\beq
\max_{\ell^t}\E[\W_t]= \frac{1}{2}\ln\left(1+\frac{\vq_\tau}{\sq_{y|x}}\right)-\frac{1}{2}\left(\sq_{x|y}-\ln\sq_{x|y}-1\right)+\frac{1}{2}\left(\vq_\tau+e^{-2\tau}\sq_{x|y}-\ln\vq_\tau-1\right),
\eeq
which, given that $\vq_\tau=1-e^{-2\tau}$ and $\sq_{x|y}=(\varsigma^{-2}_\tau+\sigma^{-2}_{y|x})^{-1}$, simplifies to $\max_{\ell^t}\E[\W_t]=\vq_\tau (1-\sq_{x|y})/2$, or, in terms of $\tau$ and $\sq_{y|x}$ only,
\beq
\max_{\ell^t}\E[\W_t]= \frac{1}{2}(1-e^{-2\tau})(1-((1-e^{-2\tau})^{-1}+\sigma^{-2}_{y|x})^{-1}).
\eeq
When $\tau\to\infty$, we recover the equilibrium result, $\E[\W_t]\leq I(X;Y)-\min_\phi\E_Y[D(P_{X|Y}(.-Y)\| G_1(.-\phi(Y)))]$, with $I(X;Y)=[\ln(1+1/\sq_{y|x})]/2$ and $\min_\phi\E_Y[D(P_{X|Y}(.-Y)\| G_1(.-\phi(Y)))]=D(G_{\sq_{x|y}}\| G_1)=(\sq_{x|y}-\ln\sq_{x|y}-1)/2$.\\


\begin{thebibliography}{10}

\bibitem{Bialek}
W~Bialek.
\newblock {\em Biophysics: Searching for Principles}.
\newblock Princeton University Press, 2013.

\bibitem{Nemenman:2011tb}
I~Nemenman.
\newblock {\em Quantitative Biology: From Molecular to Cellular Systems},
  chapter {4 Information theory and adaptation}.
\newblock CRC Press, 2012.

\bibitem{Tkacik:2011jr}
G~Tka{\v c}ik and A~M Walczak.
\newblock {Information transmission in genetic regulatory networks: a review}.
\newblock {\em Journal of Physics: Condensed Matter}, 23:153102, 2011.

\bibitem{Brennan:2012uo}
M~D Brennan, R~Cheong, and A~Levchenko.
\newblock {How information theory handles cell signaling and uncertainty}.
\newblock {\em Science}, 338:334--335, 2012.

\bibitem{Bowsher:2014cd}
CG~Bowsher and PS~Swain.
\newblock {Environmental sensing, information transfer, and cellular
  decision-making}.
\newblock {\em Current opinion in biotechnology}, 28:149--155, 2014.

\bibitem{CoverThomas91}
TM~Cover and JA~Thomas.
\newblock {\em {Elements of information theory}}.
\newblock Wiley Interscience, 1991.

\bibitem{Shannon:1948wk}
CE~Shannon.
\newblock {A mathematical theory of communication}.
\newblock {\em The Bell System Technical Journal}, 27:379, 1948.

\bibitem{Rivoire:2011ue}
O~Rivoire and S~Leibler.
\newblock {The Value of Information for Populations in Varying Environments}.
\newblock {\em Journal of Statistical Physics}, 142:1124--1166, 2011.

\bibitem{KellyJr:1956un}
JL~Kelly.
\newblock {A new interpretation of information rate}.
\newblock {\em Information Theory}, 2:185--189, 1956.

\bibitem{Barron:1988ul}
AR~Barron and TM~Cover.
\newblock {A bound on the financial value of information}.
\newblock {\em IEEE Transactions on Information Theory}, 34:1097--1100, 1988.

\bibitem{Bergstrom:2004um}
CT~Bergstrom and M~Lachmann.
\newblock {Shannon information and biological fitness}.
\newblock {\em IEEE Proc Info Theory}, pages 50--54, 2004.

\bibitem{Kussell:2005tk}
E~Kussell and S~Leibler.
\newblock {Phenotypic Diversity, Population Growth, and Information in
  Fluctuating Environments}.
\newblock {\em Science}, 309:2075--2078, 2005.

\bibitem{Taylor:2007uc}
SF~Taylor, N~Tishby, and W~Bialek.
\newblock {Information and fitness}.
\newblock arXiv:0712.4382, 2007.

\bibitem{Permuter:2011jr}
H~H Permuter, Y-H Kim, and T~Weissman.
\newblock {Interpretations of Directed Information in Portfolio Theory, Data
  Compression, and Hypothesis Testing}.
\newblock {\em IEEE Transactions on Information Theory}, 57:3248--3259, 2011.

\bibitem{Cheong:2011jp}
R~Cheong, A~Rhee, CJ~Wang, I~Nemenman, and A~Levchenko.
\newblock {Information transduction capacity of noisy biochemical signaling
  networks.}
\newblock {\em Science}, 334:354--358, 2011.

\bibitem{Rivoire:2014kta}
O~Rivoire and S~Leibler.
\newblock {A model for the generation and transmission of variations in
  evolution}.
\newblock {\em Proceedings of the National Academy of Sciences},
  111:E1940--E1949, 2014.

\bibitem{Ziv:2007cb}
E~Ziv, I~Nemenman, and C~H Wiggins.
\newblock {Optimal Signal Processing in Small Stochastic Biochemical Networks}.
\newblock {\em PLoS ONE}, 2:e1077, 2007.

\bibitem{Tkacik:2008dq}
G~Tka{\v c}ik, CG~Callan, and W~Bialek.
\newblock {Information flow and optimization in transcriptional regulation.}
\newblock {\em Proceedings of the National Academy of Sciences},
  105:12265--12270, 2008.

\bibitem{Tostevin:2009ja}
F~Tostevin and PR~ten Wolde.
\newblock {Mutual Information between Input and Output Trajectories of
  Biochemical Networks}.
\newblock {\em Physical Review Letters}, 102:218101, 2009.

\bibitem{Shannon:1949uo}
CE~Shannon.
\newblock {Communication in the presence of noise}.
\newblock {\em Proceedings of the IRE}, 37:10--21, 1949.

\bibitem{Lynch98}
M~Lynch and B~Walsh.
\newblock {\em Genetics and analysis of quantitative traits}.
\newblock Sunderland: Sinauer., 1998.

\bibitem{Kalman:1960tn}
R~Kalman.
\newblock {A new approach to linear filtering and prediction problems}.
\newblock {\em Journal of Basic Engineering}, 82:35--45, 1960.

\bibitem{Vinkler:2014vc}
DA~Vinkler, HH~Permuter, and N~Merhav.
\newblock {Analogy Between Gambling and Measurement-Based Work Extraction}.
\newblock {\em Information Theory (ISIT), IEEE International Symposium on},
  pages 1111--1115, 2014.

\bibitem{Parrondo:2015cv}
JMR Parrondo, JM~Horowitz, and T~Sagawa.
\newblock {Thermodynamics of information}.
\newblock {\em Nature Physics}, 11:131--139, 2015.

\bibitem{Haccou:1995tf}
P~Haccou and Y~Iwasa.
\newblock {Optimal mixed strategies in stochastic environments}.
\newblock {\em Theoretical Population Biology}, 47:212--243, 1995.

\bibitem{Swain:2002ww}
PS~Swain, MB~Elowitz, and ED~Siggia.
\newblock {Intrinsic and extrinsic contributions to stochasticity in gene
  expression}.
\newblock {\em Proceedings of the National Academy of Sciences},
  99:12795--12800, 2002.

\bibitem{Massey:1990vy}
J~Massey.
\newblock {Causality, feedback and directed information}.
\newblock {\em Proc. Int. Symp. Inf. Theory Applic. (ISITA-90)}, pages
  303--305., 1990.

\bibitem{Schreiber:2000wo}
T~Schreiber.
\newblock {Measuring information transfer}.
\newblock {\em Physical Review Letters}, 85:461, 2000.

\bibitem{Amblard:2013gv}
P-O Amblard and O~Michel.
\newblock {The Relation between Granger Causality and Directed Information
  Theory: A Review}.
\newblock {\em Entropy}, 15:113--143, 2013.

\bibitem{Leff02}
H~Leff and A~F Rex.
\newblock {\em Maxwell's Demon 2 Entropy, Classical and Quantum Information,
  Computing}.
\newblock CRC Press, 2002.

\bibitem{Sagawa:2008ev}
T~Sagawa and M~Ueda.
\newblock {Second Law of Thermodynamics with Discrete Quantum Feedback
  Control}.
\newblock {\em Physical Review Letters}, 100:080403, 2008.

\bibitem{Esposito:2011he}
M~Esposito and C~Van~den Broeck.
\newblock {Second law and Landauer principle far from equilibrium}.
\newblock {\em Europhysics Letters}, 95:40004, 2011.

\bibitem{Horowitz:2011ij}
JM~Horowitz and JMR Parrondo.
\newblock {Designing optimal discrete-feedback thermodynamic engines}.
\newblock {\em New Journal of Physics}, 13:123019, 2011.

\bibitem{Abreu:2011ft}
D~Abreu and U~Seifert.
\newblock {Extracting work from a single heat bath through feedback}.
\newblock {\em Europhysics Letters}, 94:10001, 2011.

\bibitem{Bauer:2012ks}
M~Bauer, D~Abreu, and U~Seifert.
\newblock {Efficiency of a Brownian information machine}.
\newblock {\em Journal of Physics A: Mathematical and Theoretical}, 45:162001,
  2012.

\bibitem{Csiszar:2008wd}
I~Csisz{\'a}r.
\newblock {Axiomatic Characterizations of Information Measures}.
\newblock {\em Entropy}, 10:261--273, 2008.

\bibitem{Horowitz:2014ev}
JM~Horowitz and H~Sandberg.
\newblock {Second-law-like inequalities with information and their
  interpretations}.
\newblock {\em New Journal of Physics}, 16:125007, 2014.

\bibitem{Cover:1996we}
TM~Cover and E~Ordentlich.
\newblock {Universal portfolios with side information}.
\newblock {\em IEEE transactions on information theory}, 42:348--363, 1996.

\bibitem{Mustonen:2010ig}
V~Mustonen and M~Lassig.
\newblock {Fitness flux and ubiquity of adaptive evolution}.
\newblock {\em Proceedings of the National Academy of Sciences},
  107:4248--4253, 2010.

\bibitem{Lan:2012in}
G~Lan, P~Sartori, S~Neumann, V~Sourjik, and Y~Tu.
\newblock {The energy--speed--accuracy trade-off in sensory adaptation}.
\newblock {\em Nature Physics}, 8:422--428, 2012.

\bibitem{Lang:2014ir}
AH~Lang, CK~Fisher, T~Mora, and P~Mehta.
\newblock {Thermodynamics of Statistical Inference by Cells}.
\newblock {\em Physical Review Letters}, 113:148103, 2014.

\bibitem{England:2013ed}
JL~England.
\newblock {Statistical physics of self-replication}.
\newblock {\em The Journal of Chemical Physics}, 139:121923, 2013.

\bibitem{Francois:2014hg}
P~Fran{\c c}ois.
\newblock {Evolving phenotypic networks in silico}.
\newblock {\em Seminars in Cell and Developmental Biology}, 35:90--97, 2014.

\bibitem{Hemery:2015ei}
M~Hemery and O~Rivoire.
\newblock {Evolution of sparsity and modularity in a model of protein
  allostery}.
\newblock {\em Phys Rev E}, 91:042704, 2015.

\end{thebibliography}
\end{document}